\documentclass[universe,review,accept,pdftex]{Definitions/mdpi} 

\usepackage{xcolor}
\usepackage{subfigure}
\usepackage{graphicx}
\usepackage{amssymb}
\usepackage{soul}
\usepackage{natbib}

\setcitestyle{citestyle=authoryear,open={(},close={)}}

\usepackage{booktabs,caption}
\usepackage[flushleft]{threeparttable}
\def\stacksymbols #1#2#3#4{\def\theguybelow{#2}
        \def\verticalposition{\lower#3pt}
        \def\spacingwithinsymbol{\baselineskip0pt\lineskip#4pt}
        \mathrel{\mathpalette\intermediary#1}}
\def\intermediary #1#2{\verticalposition\vbox{\spacingwithinsymbol
        \everycr={}\tabskip0pt
        \halign{$\mathsurround0pt#1\hfil##\hfil$\crcr#2\crcr
                \theguybelow\crcr}}}
\def\lta{\stacksymbols{<}{\sim}{2.5}{.2}}
\def\gta{\stacksymbols{>}{\sim}{2.5}{.2}}

\newcommand{\Mbh}{M$_\bullet$}
\def\sigmaT{$\sigma_{\rm v}$--T$_{\rm x}$}
\def\Tsigma{T$_{\rm x}$--$\sigma_{\rm v}$}
\def\Lsigma{L$_{\rm x}$--$\sigma_{\rm v}$}
\def\LMtwo{L$_{\rm x}$--M$_{\rm 200}$}
\def\MstarM{M$_{\rm stars}$--M}
\def\LM{L$_{\rm x}$--M}
\def\LT{L$_{\rm x}$--T$_{\rm x}$}
\def\MT{M--T$_{\rm x}$}
\def\MY{M--Y$_{\rm x}$}

\def\tx{\rm T_x}
\def\R500{R$_{500}$}
\def\r500{{\rm R_{500}}}

\firstpage{1} 
\pubvolume{7}
\issuenum{5}
\articlenumber{139}
\pubyear{2021}
\copyrightyear{2021}
\history{Received: 31 March 2021; Accepted: 28 April 2021; Published: 10 May 2021}

\Title{Scaling Properties of Galaxy Groups}

\Author{Lorenzo Lovisari$^{1,2,\ddagger}$\orcidA{}, Stefano Ettori$^{1,3,\ddagger}$\orcidB{}, Massimo Gaspari$^{1,4,\ddagger}$\orcidC{}, and Paul A. Giles$^{5,\ddagger}$\orcidD{}}

\AuthorNames{Lorenzo Lovisari, Stefano Ettori, Massimo Gaspari, and Paul Giles}

\address{%
$^{1}$ \quad INAF - Osservatorio di Astrofisica e Scienza dello Spazio di Bologna, via Piero Gobetti 93/3, 40129 Bologna, Italia; lorenzo.lovisari@inaf.it\\
$^{2}$ \quad Center for Astrophysics $|$ Harvard $\&$ Smithsonian, 60 Garden Street, Cambridge, MA 02138, USA;\\
$^{3}$ \quad INFN, Sezione di Bologna, viale Berti Pichat 6/2, 40127 Bologna, Italia; \\
$^{4}$ \quad Department of Astrophysical Sciences, Princeton University, 4 Ivy Lane, Princeton, NJ 08544, USA;\\
$^{5}$ \quad Department of Physics and Astronomy, University of Sussex, Falmer, Brighton BN1 9QH, UK.
}

\firstnote{All authors contributed equally to this work.}

\abstract{
Galaxy groups and poor clusters are more common than rich clusters, and host the largest fraction of matter content in the Universe. Hence, their studies are key to understand the gravitational and thermal evolution of the bulk of the cosmic matter. 
Moreover, because of their shallower gravitational potential, galaxy groups are systems where non-gravitational processes (e.g., cooling, AGN feedback, star formation) are expected to have a higher impact on the distribution of baryons, and on the general physical properties, than in more massive objects, inducing systematic departures from the expected scaling relations. \\
Despite their paramount importance from the astrophysical and cosmological point of view, the challenges in their detection have limited the studies of galaxy groups.  
Upcoming large surveys will change this picture, reassigning to galaxy groups their central role in studying the structure formation and evolution in the Universe, and in measuring the cosmic baryonic content. \\
Here, we review the recent literature on various scaling relations between X-ray and optical properties of these systems, focusing on the observational measurements, and the progress in our understanding of the deviations from the self-similar expectations on groups' scales. 
We discuss some of the sources of these deviations, and how feedback from supernovae and/or AGNs impacts the general properties and the reconstructed scaling laws.
Finally, we discuss future prospects in the study of galaxy groups. }


\begin{document}
\section{Introduction} \label{s:intro}
Following the hierarchical scenario of structure formation, galaxy systems form through episodic mergers of small mass units. The less massive ones (often referred as groups), are the building blocks for the most massive ones (clusters), and trace the filamentary components of the large scale structure (e.g., \citealt{eke04}). However, the distinction between groups and clusters is quite loose and no universal definition exists in literature. Also, because the halo mass function is continuous, a naive starting point would be to not single out the low-mass end objects. Nonetheless, these poor systems have some notable differences (e.g., lack of dominance of the gas mass over the stellar/galactic component; \citealt{giodini09}) with respect to their more massive counterpart and they cannot be simply considered their scaled-down versions. 

A conventional "rule of thumb" definition is to label systems of less than 50 galaxies as groups and above as clusters. More in general, galaxy groups have been broadly classified into three main classes on the basis of their optical and physical characteristics: poor/loose groups, compact groups, and fossil groups (e.g., \citealt{2007AN....328..699E}).  Poor/loose groups are aggregate of galaxies with a space density of $\sim$10$^{-5}$ Mpc$^{-3}$ (e.g., \citealt{1987MNRAS.225..505N}). Compact groups are small and relatively isolated systems of typically 4-10 galaxies with a space density of $\sim$10$^{-6}$ Mpc$^{-3}$ (e.g., \citealt{1982ApJ...255..382H}). Fossil groups are objects dominated by a single bright elliptical galaxy (a formal definition is provided in \citealt{2003MNRAS.343..627J}). Early studies (e.g., \citealt{helsdon00}) showed that subsamples of loose and compact groups share the same scaling relations. Thus, in this review, we do not make distinction between poor/loose and compact groups, and hereafter we simply refer to them as galaxy groups. The properties of fossil groups are instead discussed in the companion review by \cite{2021Univ....7..132A}. 
However, since the optical properties are not always available, a threshold of M$\sim$10$^{14}$M$_{\odot}$, corresponding to a temperature of 2-3 keV, is also often used to classify these systems. We will show later that this threshold roughly corresponds to the temperature for which there is a significant change in the X-ray emissivity. 

Despite the crucial role played by groups in cosmic structure formation and evolution, they have received less attention compared to massive clusters. One of the reasons is that typical groups contain only a few bright galaxies in their inner regions, making very difficult to detect them in optical with a relatively good confidence. A much easier method of detecting them is to study the X-ray emission from the hot intragroup medium (IGrM). The detection of hot plasma carries witness that galaxy groups (and clusters) are not simple conglomerate of galaxies put together by projection effects, but real physical systems which are undergoing some degree of virialization. 
Galaxy groups often show lower and flatter X-ray surface brightness than clusters (e.g., \citealt{Ponman:1999}, \citealt{Sanderson2003}). Therefore, the physical properties of the gas derived for galaxy groups are presumably less robust than the properties derived for galaxy clusters. Nonetheless, they represent a more common environment because the mass function of virialized systems, which describes the number density of clusters above a threshold mass M, is higher at lower masses (with a factor of $\sim$30/210/1500 more objects in the mass range M$_{500}= 10^{13}$ M$_{\odot}-{\rm M}_1$ than in M$_{500} > {\rm M}_1$, and M$_1 = 1/2/5 \times 10^{14} {\rm M}_{\odot}$ at $z=0$; see, e.g., \citealt{despali16}).  Hence, the detection and  characterization of galaxy groups is especially important for astrophysical and cosmological studies. 

\subsection{Galaxy groups and astrophysics}
\label{sec:groupastro}
Galaxy groups cover the intermediate mass range between large elliptical galaxies and galaxy clusters and contain the bulk of all galaxies and baryonic matter in the local Universe (e.g., \citealt{Tully1987,1998ApJ...503..518F}, \citealt{eke04}). Because of that, they are crucial for understanding the effects of the local environment on galaxy formation and evolution processes. Moreover, the feedback from supernovae (SNe) and supermassive black holes (SMBHs) is expected to alter significantly the properties of these systems being the energy input associated with these sources comparable to the binding energies of groups (e.g., \citealt{Brighenti:2002,McCarthy:2010,Gaspari:2012b}).  However, the relative contributions of the different feedback processes is still a matter of debate and it will constitute a major subject of research for the next decade. These factors make galaxy groups great laboratories to  understand the complex baryonic physics involved, and to study the differences with their massive counterpart. For instance, we know that the fraction of strong cool-cores (CC; i.e., systems with a central cooling time $t_{\rm cool}<$ 1 Gyr, as described in \citealt{2010A&A...513A..37H}), weak cool-cores (1 $<t_{\rm cool}<$ 7.7 Gyr), and non cool-cores (NCC; $t_{\rm cool}>$ 7.7 Gyr) objects at the group scale are similar to those in galaxy clusters (\citealt{Bharad14}). However, \citet{Ewan2017} found that the CC fraction increases dramatically when the samples are restricted to low-temperature systems (i.e., kT$<$1.5 keV) showing a correlation between system temperature and CC status. \citet{Bharad14} also found that brightest group galaxies have a higher stellar mass than brightest cluster galaxies, suggesting that there is less gas available to feed the SMBHs. Recent results suggest that the IGrM and intracluster medium (ICM) are also providing a source of gas which feeds and grows the central SMBHs, in particular leading to novel scaling relations between the SMBH mass and the X-ray properties of their host gaseous halos (e.g., \citealt{Bogdan2018,Gaspari:2019,2019MNRAS.488L.134L}).
These findings imply an interplay between the feedback mechanisms connected with the SMBHs and the macro-scale halos, which could explain some features of cosmological simulations driving a relative break of the L$_{\rm x}$--T$_{\rm x}$ and L$_{\rm x}$--M relations at low temperatures 
(e.g., \citealt{2007MNRAS.380..877S,Puchwein:2008,2010MNRAS.401.1670F,McCarthy:2010,2014MNRAS.441.1270L}). This deviation is often attributed to active galactic nucleus (AGN) feedback (e.g., \citealt{2014MNRAS.438..195P,Gaspari:2014_scalings,Truong:2018}). 

The properties described above have an important effect on the correlation between different physical quantities. For instance, it is well established that CC and NCC objects populate different regions of the X-ray luminosity space of any scaling relations (e.g., \citealt{Markevitch1998,Pratt09,2011A&A...532A.133M,Bharad2015,2016MNRAS.463.3582M,LL20}). Therefore, a change in the fraction of CC/NCC systems as a function of the temperature (mass) will have an impact to the slope, normalization, and scatter of the observed scaling relations. Hence, it is crucial to have a full coverage for the whole sample to minimize the systematic errors due to the incompleteness.  In fact, if all the missing objects happen to belong to one of the subsamples (e.g., NCC), the normalization (and the scatter) of the studied scaling relations will be wrong. Moreover, the CC/NCC fraction of systems in a sample depends on the selection function and may not be representative of the underlying population. For instance, X-ray selected samples are known to be biased toward centrally peaked and relaxed systems, in particular in the low-mass regime (\citealt{2011A&A...526A..79E}).  
In fact, recent results by \citet{Ewan2017}, who analyzed a sample of optically selected groups, show that $\sim$20\% of X-ray bright groups (probably the most disturbed ones, or with no concentrated CC) in the local Universe may have been missed. Thus, the scaling relations of galaxy groups (and clusters) are the result of the various processes that govern the formation and evolution of these systems making them ideal targets for studying the effect of the interplay between galaxy evolution, the development of the IGrM, and feedback.

\subsection{Galaxy groups and cosmology}
\label{sec:groupcosmo}
Clusters of galaxies have proven to be remarkably effective probes of cosmology (e.g., \citealt{Reiprich2002, Allen08, 2009ApJ...692.1060V, Rozo10, Mantz15, dehaan16, Planck16, bohr17, 2017MNRAS.471.1370S, 2018A&A...620A..10P, Bocquet2019}.  
However, since galaxy groups represent a large fraction of the number density of virialized systems, their impact might be relevant (in particular, on the reconstruction of the halo mass function).
For instance, recent results of the Dark Energy Survey (DES) collaboration show that the $\sigma_{8}$--$\Omega_{\rm M}$ posteriors have a 5.6$\sigma$ tension with \textit{Planck} CMB results, and a 2.4$\sigma$ tension with galaxy clustering and cosmic shear results (\citealt{DESY13x2pt}). The cause of this tension is thought to reside at the low-mass (low richness) end of the cluster population, specifically, clusters with a richness of $\lambda<30$ (corresponding to $\sim$10$^{14}$ M$_{\odot}$). 
The removal of low richness systems from the analysis significantly reduces the tension with comparative cosmological probes. 
However, various tests undertaken in \citet{DESY1cosmo} suggest that the discrepancy is probably due to the modelling of the weak lensing signal rather than the group and cluster abundance. The mass calibration for the DESY1 analysis is based upon a stacked weak lensing analysis, through application of the weak lensing--richness relation (\citealt{mcclintock19}).  This relation is derived over the full richness range, which would not account for any deviations at the low-mass end. 
Furthermore, since the mass analysis relies on stacked quantities, information on scatter in mass with richness is lost and must be informed from external relations.  In the case of the DESY1 analysis, the mass scatter information is inferred from the temperature--richness relation using X-ray data (\citealt{farahi19}).  This scatter is assumed constant with richness, which again, could evolve as a function of richness.  The investigation of these effects will become of critical importance as the low-mass end of the mass scales are increasingly probed by future surveys
(e.g., those constructed from the Legacy Survey of Space and Time undertaken by the {\it Vera C. Rubin Observatory}).

Excluding low-mass systems significantly reduces the cosmological parameter constraints. Thus, despite the important complications present at the group scale, it is becoming generally appreciated that galaxy groups should be included in the cosmological analysis. In order to use them to constrain the cosmological parameters we need a good knowledge of the selection function to properly correct for the incompleteness, otherwise studies employing the cluster mass function may find lower $\Omega_{\rm M}$ and/or $\sigma_8$ values than the true values. This scenario is supported by the finding of \citet{2017MNRAS.471.1370S}, who showed how the increasing incompleteness of parent samples in the low-mass regime together with a steeper L$_{\rm x}$--M relation observed for groups, can lead to biased cosmological parameters. It is worth noticing that, if a large fraction of galaxy systems is missed, than the tension between cluster counts and primary CMB cosmological constraints may become less severe. 

Most of the upcoming large surveys will push the measurements down to the low-mass regime. Thus, to fully exploit the future datasets to constrain the cosmological parameters, we need to properly characterize the properties of galaxy groups and the differences with galaxy clusters, accounting for the different selection effects, and estimating the amplitude of the various biases. 

\subsection{This review}
In this work, we present an overview of the most recent studies on scaling relations between a number of integrated observed quantities of galaxy groups, and complement/update the previous reviews in the field by, e.g.,  \citet{Mulchaey2000} and \citet{Sun2012}. The review is organized as follows. In Section \ref{s:sss}, we derive the self-similar X-ray scaling relations and overview the observed deviations. In Section \ref{s:multi}, we discuss the relations between X-ray and optical properties. In Section \ref{s:coev}, we discuss the relation between the SMBH mass and the global group quantities. In Section \ref{future}, we shortly discuss the most relevant upcoming missions and their expected contribution to the field. In Section \ref{sect:remarks} we provide our final remarks.

\section{X-ray scaling relations} \label{s:sss}

\subsection{Theoretical expectations} \label{selfsimilar}

The X-ray scaling relations for galaxy systems were derived by \citet{Kaiser1986}, based on the simple assumption that the thermodynamic properties of the ICM are only determined by gravity (i.e., gas just follow the dark matter collapse). Since gravity is scale free, this model predicts that objects of different sizes are the scaled version of each other. For that reason, this model is often referred as self-similar, and the derivation of the predicted relations has been extensively covered in the literature (e.g., \citealt{1996ApJ...469..480K,1998ApJ...495...80B,Voit2005,Maughan2006,2008SSRv..134..269B,Bohringer2012,Ettori2013,Giodini2013,Maughan2014,Ettori2015,Ettori2020}).  Here, we only provide a brief review of the standard derivation of the self-similar scaling relations for massive systems, and then extend them, when necessary, to the low-mass regime where gas physics is playing a significant role.

In the self-similar scenario, two galaxy systems which have formed at the same time have the same mean density. Hence, 
\begin{equation}
\frac{{\rm M}_{\Delta_z}}{{\rm R}^3_{\Delta_z}}={\rm constant}
\end{equation}
where \textrm{M$_{\Delta_z}$} is the mass contained within the radius \textrm{R$_{\Delta_z}$}, encompassing a mean density ${\Delta_z}$ times the critical density of the Universe  $\rho_c(z)$, so that M$_{\Delta_z}\propto\rho_c(z){\Delta_z}{\rm R}^3_{\Delta_z}$. The critical density of the Universe scales with redshift as $\rho_c(z)$=$\rho_{c(z=0)}$E$^2(z)$, where E$(z)$=H$_z$/H$_0$ describes the evolution of the Hubble parameter with redshift $z$.

During the gravitational collapse the gas density increases, and a shock propagates outward from the cluster center and heats the gas.  After the passage of the shock, IGrM and ICM can be considered in hydrostatic equilibrium, so the temperature T$_{\rm x}$ provides an estimate of the gravitational potential well (i.e., T$_{\rm x}\propto$ GM$_{\Delta_z}$/R$_{\Delta_z}\propto {\rm R}_{\Delta_z}^2$), and therefore of the total mass of the cluster:
\begin{equation}\label{eq:MT}
{\rm M}_{\Delta_z}\propto {\rm T_{\rm x}}^{3/2}.
\end{equation}
In the self-similar scenario, where the gas fraction, f$_{\rm g}$, of galaxy groups and clusters is universal, one expects for the total gas mass, M$_{\rm g}$, a similar dependence on the gas temperature: M$_{\rm g}\propto {\rm T_{\rm x}}^{3/2}$. 

The hot gas in galaxy systems is typically described as an optically thin plasma in collisional ionisation equilibrium. Its X-ray emissivity (i.e., the energy emitted per time and volume) is equal to 
\begin{equation}
\epsilon= {\rm n_e} \, {\rm n_p} \, \Lambda({\rm T_x},{\rm Z}_{\odot}),
\end{equation}
where ${\rm n_e}$ and ${\rm n_p}$ are the number densities of electrons and protons, respectively, that are related to the gas mass density $\rho_{\rm g}$ through the relation $\rho_{\rm g}$=$\mu {\rm m_p} ({\rm n_e}+{\rm n_p})$, $\mu$ is the mean molecular weight ($\sim$0.6 for a plasma with solar abundance), ${\rm m_p}$ is the proton mass, and $\Lambda({\rm T_x},{\rm Z}_{\odot})$ is the cooling function which depends on the mechanism of the emission\footnote{Three main processes contribute to the X-ray emission: thermal bremsstrahlung (due to the deflection of a free electron by the electric field of a ion), recombination (due to the capture of an electron by an ion), and two-photon decay (due to the changing of the quantum level of an electron in an ion). See details in the reviews from, e.g., \citet{Sarazin1986}, \citet{2006PhR...427....1P}, \citet{2008SSRv..134..155K}, and \citet{2010A&ARv..18..127B}.} and on the considered energy window. 
At high temperatures (i.e., kT$>$3 keV) the main mechanism of emission is thermal bremsstrahlung and the cooling function in the full energy band mainly depends only on T$_{\rm x}$ (i.e., $\Lambda({\rm T_x},{\rm Z}_{\odot})\propto$T$_{\rm x}^{1/2}$). Thus, for sufficiently massive systems the bolometric X-ray luminosity (i.e., 0.01-100 keV band) is given by  
\begin{equation}\label{eq:LT}
\textrm{L$_{\rm x,bol}\propto\int \epsilon$ dV $\propto$ n$_{\rm p}^2$ T$_{\rm x}^{1/2}$R$^{3}\propto$ f$_{\rm g}^2$ T$_{\rm x}^2\propto$ T$_{\rm x}^2$}
\end{equation}
with the last scaling obtained assuming a constant gas fraction as predicted by the self-similar scenario. By combining Equations \ref{eq:MT} and \ref{eq:LT} one obtains the well known relation between bolometric luminosity and total mass (i.e., L$_{\rm x,bol}\propto$M$^{4/3}$).

\begin{figure}[t]
\centering
\includegraphics[width=0.85\textwidth]{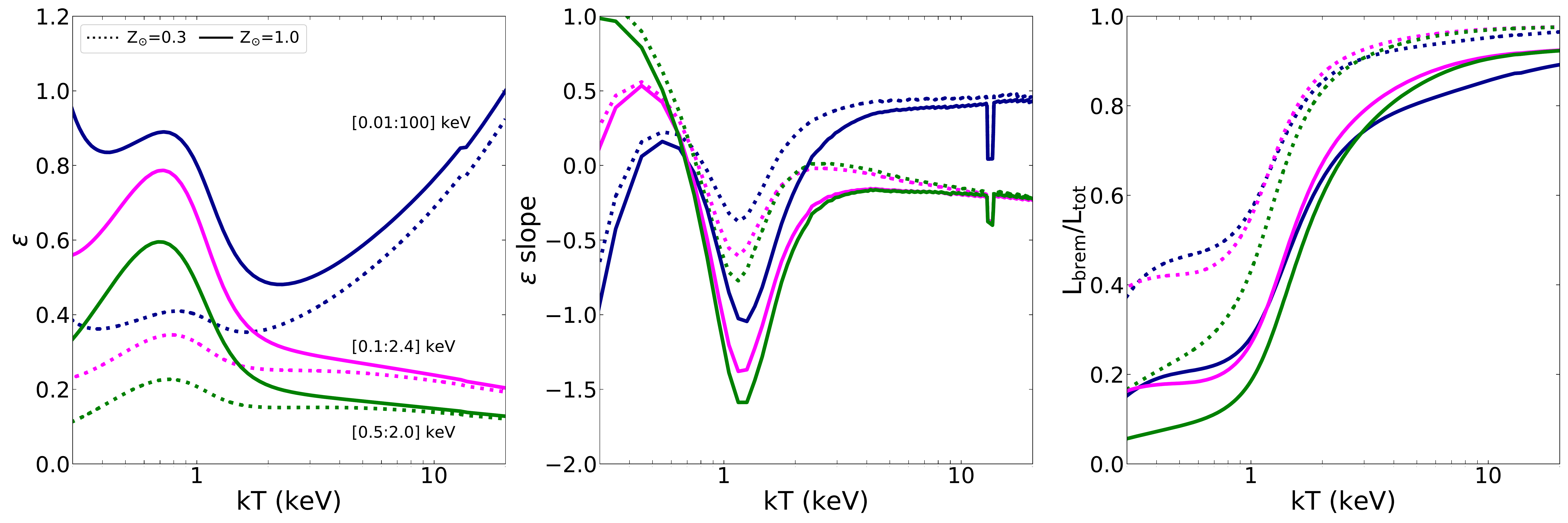}
\caption{({\it \textbf{left panel}}): total X-ray emissivity as function of the plasma temperature in different energy bands (bolometric in blue, 0.1--2.4 keV in magenta, and 0.5--2 keV in green). The curves are calculated using an APEC (\citealt{APEC2001}) model (v3.0.9) in XSPEC (\citealt{xspec1996}) for two different values of metallicity: 1.0 (solid lines) and 0.3 times the solar abundance as in \citet{Asplund2009}. All curves are normalized to the bolometric emissivity at k$_{\rm B}$T$_{\rm x}$ = 20 keV with \mbox{Z = 1\,Z$_{\odot}$}. ({\it \textbf{middle panel}}): the emissivity slope as a function of temperature showing the impact of the different $Z_{\odot}$ and T$_{\rm x}$ in the low-temperature regime. ({\it \textbf{right panel}}): bremsstrahlung emission fraction (L$_{\rm brem}$/L$_{\rm tot}$) as a function of the temperature, illustrating the increasing contribution of line emission to the total luminosity for low-temperature plasmas.}
\label{fig:emissivity}
\end{figure} 

In the literature the X-ray luminosities are also often provided in soft energy bands (e.g., 0.1-2.4 or 0.5-2 keV), more representative of the bandpass covered by current (and past) X-ray facilities used for the study of groups and clusters.  In Figure \ref{fig:emissivity} ({\it left panel}), we show that for massive systems with typical cluster abundance the X-ray emissivity in soft band is almost independent of the system temperature (e.g., for Z=0.3Z$_{\odot}$ the change in $\epsilon$ between 3 and 10 keV plasmas in the 0.5-2 keV band is $<$10\% for given emission measure), so that L$_{\rm x,soft}\propto$T$_{\rm x}^{3/2}$, and hence using Equation \ref{eq:MT}, L$_{\rm x,soft}\propto$M.

However, the gas fraction is not constant, with a difference of almost a factor of two between groups and clusters (e.g., \citealt{2006ApJ...640..691V,gonzalez2007,2007ApJ...669..158G}, \citealt{Pratt09,2010ApJ...719..119D,gonzalez13,LL15,2016A&A...592A..12E}; see also the companion reviews by \citealt{2021Univ....7..142E} and \citealt{universe7070209}). Moreover, at low temperatures, line cooling becomes very important, and the emissivity (both in soft and bolometric bands) becomes strongly abundance (Z$_{\odot}$) and temperature dependent. In Figure \ref{fig:emissivity} ({\it left} and {\it middle panels}) we show the dependence of the emissivity on the temperature and metallicity for widely used energy bands for scaling relations,  clearly showing that a simple scaling cannot be derived. In Table \ref{table:em}, we provide the dependence for a set of interesting cases. 

The complexity of the emissivity function in the low-temperature regime may lead to a wrong interpretation of the results of scaling relation studies. In fact, it is conventional to compare the slopes of the scaling relations obtained with sample of groups to the self-similar predictions derived for massive clusters. However, if there is no feedback (i.e., the relations follow the self-similar predictions), then the L$_{\rm x}$--T$_{\rm x}$ and L$_{\rm x}$--M relations should flatten at low temperatures and masses. Thus, without accounting for the increasing contribution of the line emission in the low-temperature regime, one could interpret the agreement between group and cluster relations such that feedback processes play a negligible role in shaping the IGrM. Thus, the impact of the feedback could be underestimated.
To visualize the contribution of line emission as function of the temperature, we follow the simple approach of \citet{Zou2016} in which we measure the luminosity (L$_{\rm tot}$) in different energy bands (i.e., bolometric, 0.1-2.4, and 0.5-2) of APEC spectra with a metal abundance of Z$_{\odot}$=1.0 (not rare at the center of galaxy groups, see companion review by \citealt{Gastaldello:2021}), and then setting Z$_{\odot}$=0 without changing any other parameters to approximate the luminosity of the pure bremsstrahlung component (L$_{\rm brem}$). We repeated the exercise for a more standard Z$_{\odot}$=0.3. The results are shown in Figure \ref{fig:emissivity} ({\it right panel}) where it is clear the significant contribution of line emission to the total luminosity in the low-temperature regime. Thus, the luminosity--temperature and luminosity--mass relations can be approximated as L$_{\rm x}\propto$ T$_{\rm x}^{1.5+\gamma}$ and  L$_{\rm x}\propto$ M$^{1+\gamma}$, where $\gamma$ is the slope of the X-ray emissivity in the considered energy band (e.g., soft or bolometric) and temperature range covered by the systems in the studied dataset (see Table \ref{table:em}). It follows that the self-similar L$_{\rm x}$--T$_{\rm x}$ and L$_{\rm x}$--M relations for galaxy groups are expected to be significantly flatter than the ones for galaxy clusters.  It is also worth noticing that even for massive systems with Z$_{\odot}$=0.3 there is a $\sim$5\% contribution from line emission. Thus, the bolometric emissivity slope is smaller than 0.5 (i.e., the value one gets from pure bremsstrahlung emission) with the net effect being that the correct self-similar expectation becomes L$_{\rm x,bol}\propto$T$_{\rm x}^{\sim1.9}$. 

\begin{table}[t!]
\caption{Emissivity dependence on T$_x$ and Z$_{\odot}$ for different temperature regimes and energy bands.}
\centering
\begin{tabular}{ccccc}
\toprule
\textbf{E band} & \textbf{T range}	& \textbf{$\epsilon$ slope (Z=0.3Z$_{\odot}$}) & \textbf{$\epsilon$ slope (Z=0.5Z$_{\odot}$}) & \textbf{$\epsilon$ slope (Z=1.0Z$_{\odot}$})\\
\midrule
bol     & 0.4-0.7   & +0.20  & +0.16  & +0.11 \\
        & 0.4-2.0   & --0.00 & --0.14 & --0.34 \\
        & 0.4-3.0   & +0.06 & --0.07 & --0.26 \\
        & 0.4-10.0  & +0.20  & +0.11  & --0.03 \\
        & 0.7-2.0   & --0.11 & --0.30 & --0.58 \\
        & 0.7-3.0   & +0.01 & --0.15 & --0.40 \\
        & 0.7-10.0  & +0.20  & +0.10  & --0.06 \\
        & 2.0-10.0  & +0.40  & +0.36  & +0.28 \\
        & 3.0-10.0  & +0.43  & +0.41  & +0.35 \\
\midrule
0.1-2.4 & 0.4-0.7   &  +0.44 & +0.42  & +0.39 \\
        & 0.4-2.0   & --0.04 & --0.19 & --0.42 \\
        & 0.4-3.0   & --0.04 & --0.18 & --0.39 \\
        & 0.4-10.0  & --0.06 & --0.16 & --0.31 \\
        & 0.7-2.0   & --0.29 & --0.52 & --0.84 \\
        & 0.7-3.0   & --0.22 & --0.40 & --0.68 \\
        & 0.7-10.0  & --0.16 & --0.27 & --0.45 \\
        & 2.0-10.0  & --0.08 & --0.12 & --0.20 \\
        & 3.0-10.0  & --0.10 & --0.12 & --0.17 \\
\midrule
0.5-2   & 0.4-0.7   &  +0.63 & +0.56  & +0.50 \\
        & 0.4-2.0   & --0.03 & --0.23 & --0.48 \\
        & 0.4-3.0   & --0.02 & --0.21 & --0.45 \\
        & 0.4-10.0  & --0.04 & --0.17 & --0.35 \\
        & 0.7-2.0   & --0.38 & --0.65 & --1.00 \\
        & 0.7-3.0   & --0.27 & --0.50 & --0.81 \\
        & 0.7-10.0  & --0.18 & --0.32 & --0.52 \\
        & 2.0-10.0  & --0.06 & --0.11 & --0.22 \\
        & 3.0-10.0  & --0.07 & --0.11 & --0.18 \\
\bottomrule
\label{table:em}
\end{tabular}
\end{table}

The abundance and temperature dependence of the X-ray emissivity at low temperatures need to be taken into account when determining the luminosities of galaxy groups. Normally, the luminosities are estimated applying a conversion factor to the observed count rates to obtain the X-ray fluxes. From Figure \ref{fig:emissivity} it is clear that this conversion factor in the low-temperature regime depends strongly on the metallicity of the system. Given the observed temperature and abundance gradients in groups (e.g., \citealt{Rasmussen2007,Sun2009,Mernier2017,LL19}; see also the companion review by \citealt{Gastaldello:2021}), a possible strategy is to use the observed profiles of temperature, abundance, and surface brightness to estimate the luminosity in each radial bin obtained during the spectral analysis (e.g., \citealt{Sun2012}, \citealt{LL15}). \citet{Sun2012} pointed-out that although the average luminosities (soft-band or bolometric) only change by $\sim$5\% when the overall values of temperature and abundance are used in the conversion instead of the profiles, the scatter increases by 10-15\%. This is an important point to keep in mind when using survey data (e.g., ROSAT, eROSITA) for which simple assumptions like isothermality and single overall abundance are chosen to obtain an estimate of the luminosity.  

\begin{figure}[h]
\centering
\includegraphics[width=0.85\textwidth]{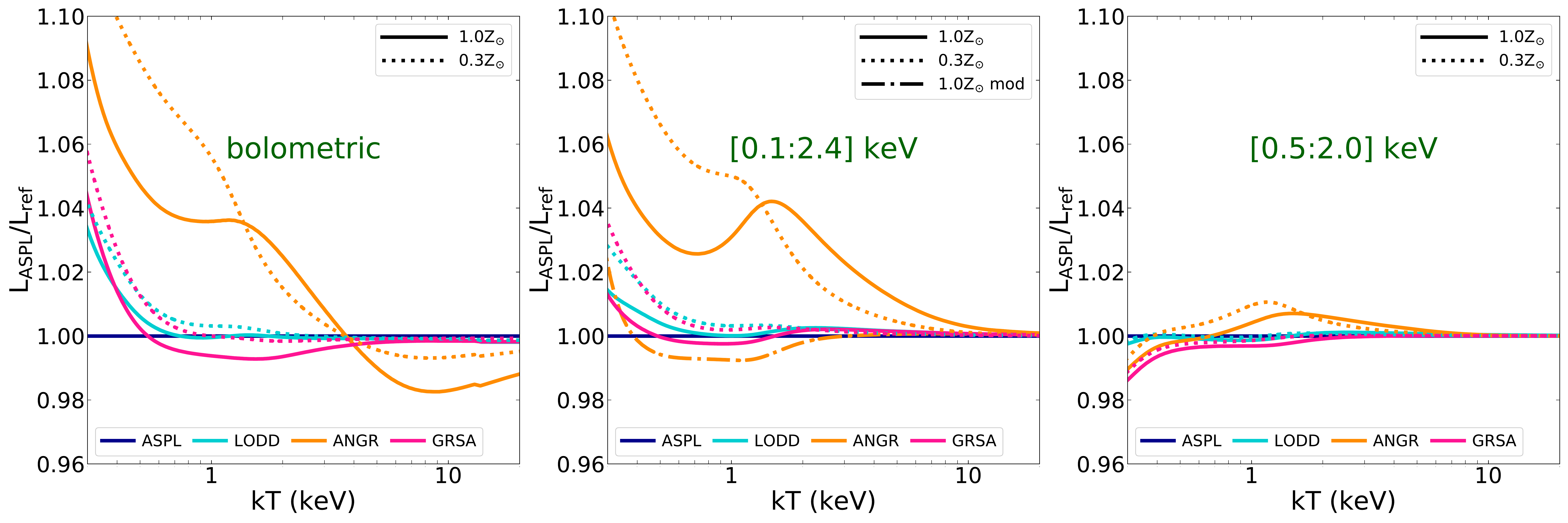}
\includegraphics[width=0.85\textwidth]{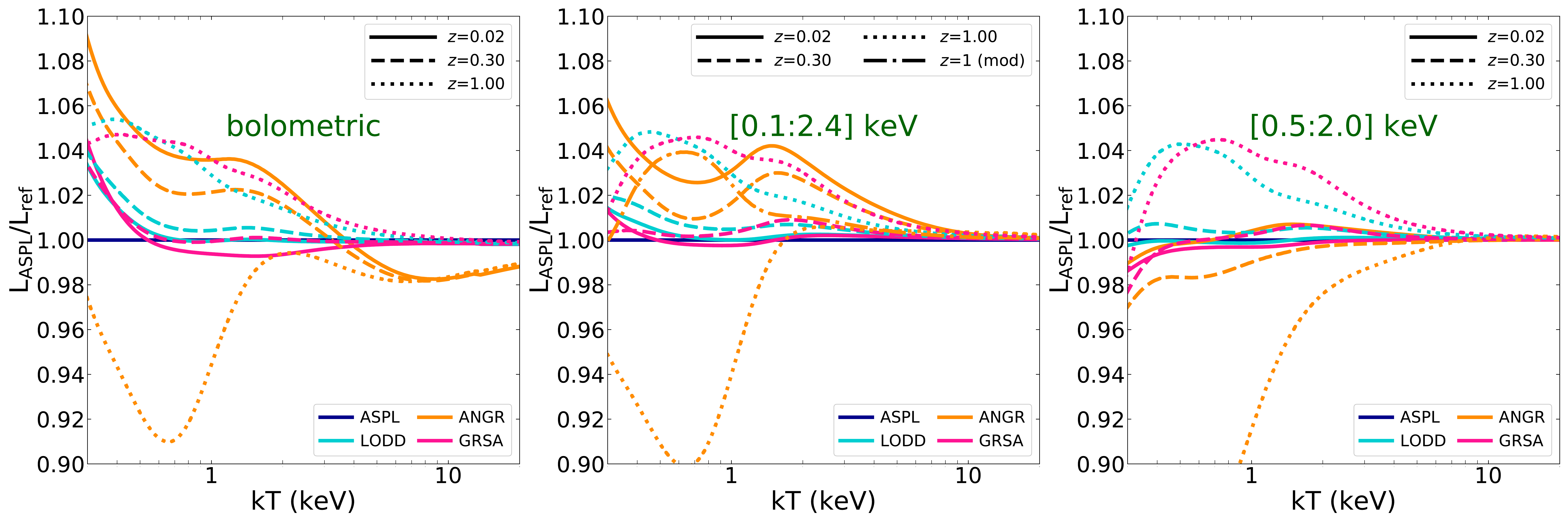}
\caption{Ratio between the rest-frame L$_{\rm x}$ ({\it\textbf{ left panels}:} bolometric, {\it \textbf{middle panels}}: 0.1--2.4 keV band, {\it \textbf{right panels}}: 0.5--2 keV band) derived using the abundances as in \citet{Asplund2009} and L$_{\rm x}$ obtained with the abundances of \citet[turquoise]{Lodders2009}, \citet[orange]{Anders1989}, and \citet[pink]{Grevesse1998}, respectively. In the {\it \textbf{top panels}} we show the results for systems at $z$=0.02 with a metallicity of 1.0 (solid lines) or 0.3 times the solar abundance. The dashdot line in the {\it \textbf{middle panels}} refers to the simulations with a modified table of ANGR in which the Fe abundance was set equal to the value of ASPL, showing that indeed most of the differences arise from the significant discrepancy in Fe between ANGR and the other tables. 
The 0.5--2 keV band (i.e., the band used to rescale the APEC normalization) provide the best agreement between the different tables. In the {\it \textbf{bottom panels}}, we show the impact on L$_{\rm x}$ of changing abundance table for systems at different redshifts: $z$=0.02 (solid line), $z$=0.3 (dashed), and $z$=1 (dotted, dashdotted). The simulations were performed with Z=1Z$_{\odot}$. The plot shows how the differences increases with $z$, although the effect is generally smaller than $\sim$5\% unless very high $z$ are considered. The residual difference between ASPL and the modified ANGR table for high $z$ objects is due to the  differences in elements other than Fe (e.g., C, N, O, Ne, Mg, Si, S).        
}
\label{fig:emissabund}
\end{figure} 

The dependence of the cooling function on the metallicity also implies that the use of different abundance tables can lead to different estimates of the rest-frame X-ray luminosities. Typically, one recovers the source count rate within a given radius from the surface brightness profile, and then obtain the X-ray flux by setting the normalization of a thermal model (with proper temperature and metallicity) to match the observed count rate. However, the shape of the thermal model (which depends only on the abundance for a given temperature and column density) can diverge at lower and higher energies than the ones used to derive the surface brightness. To visualize the impact, we ran a set of simulations in which the normalization of the thermal model for systems at $z$=0.02 (i.e., median redshift of the current local group samples, see Table \ref{slsumary}) was set in order to match a count rate of 1 count/sec in the 0.5-2 keV energy band (i.e., the bandpass where many X-ray facilities have most of their effective area, and often used to derive the surface-brightness profiles) for each abundance table. Then, we estimated the luminosity in different energy bands. In Figure \ref{fig:emissabund} ({\it top panels}) we show the impact on the estimated luminosity as function of the system temperature and common abundance tables. There is a very good agreement in the 0.5-2 keV band luminosity, regardless of the abundance table used for the analysis. Instead, small differences (i.e., in the order of a few percent) in the 0.1-2.4 keV band and bolometric luminosities arise for low temperature systems (i.e., kT$\lesssim$1 keV) when the abundances of \citet[GRSA]{Grevesse1998}, \citet[ASPL]{Asplund2009}, or \citet[LODD]{Lodders2009} are used. The disagreement is much more significant (i.e., up to $\sim$10\%) when the luminosities are estimated with the abundance table by \citet[ANGR]{Anders1989}. Most of the differences are due to the much higher Fe abundance in ANGR  with respect to the other tables investigated here. When the Fe abundance of ANGR is set to the value of ASPL (leaving unchanged all the other ANGR abundances) the estimated luminosities are in much better agreement (see the dashdot lines in the {\it middle panels} of Figure \ref{fig:emissabund}). The reason for the differences highlighted in Figure \ref{fig:emissabund} is that by switching the abundance table we change the emissivity and the relative contribution of the line emission with respect to the bremsstrahlung emission (see Figure \ref{fig:emissivity}). The difference between the rest-frame luminosity estimated with one or another table tends to increase at higher redshifts (see {\it bottom panels} of Figure \ref{fig:emissabund}). However, unless very high redshifts are considered, the effect is usually smaller than a few percent. In general, the soft-bands (in particular the 0.5-2 keV band) are the ones showing a smaller impact on the estimated luminosity by switching abundance table and should be preferred for galaxy groups studies. Although, in most cases the effect is relatively small, it can lead to systematic effects  and should be kept in mind when comparing independent literature results.   

Another very useful quantity to describe the IGrM and ICM is the entropy which is generated during the hierarchical assembly process. In X–ray studies of galaxy groups and clusters, the entropy is usually defined as 
\begin{equation}
\textrm{K = k$_{\rm B}$T$_{\rm x}$ n$_{\rm e}^{-2/3}$}    
\end{equation}
where k$_{\rm B}$ is the Boltzmann constant. Entropy is conserved during adiabatic processes and it is only modified by processes changing the physical characteristics of the gas. Entropy increases when heat energy is introduced and decreases when radiative cooling carries heat energy away (e.g.,  \citealt{Voit2005}), keeping a record of the energy injection and dissipation in the intracluster gas. Thus, entropy measurements provide a useful tool for our understanding of the thermodynamic history of galaxy groups and clusters. Gas entropy in galaxy groups shows a significant excess to that achievable by pure gravitational collapse  (e.g., \citealt{Ponman:1999,2000MNRAS.315..689L,Ponman2003,Finoguenov2007}, \citealt{Sun2009,Johnson2009}, \citealt{2014MNRAS.438.2341P}), indicating a  substantial IGrM heating often ascribed to non-gravitational processes. In fact, due to the shallower potential well of the small systems, it is expected that the energy released by past star formation and AGN activities leaves a clear imprint on the thermodynamic properties of IGrM and ICM (see companion reviews by \citealt{2021Univ....7..142E} and \citealt{universe7070209}). An effect that can be seen in both the integrated properties (i.e., in the scaling relations) and in the shape of the entropy profiles which are expected to follow K$\propto$R$^{1.1}$.

\subsection{Observed scaling relations}

\begin{figure}[t]
\centering
\includegraphics[width=\textwidth]{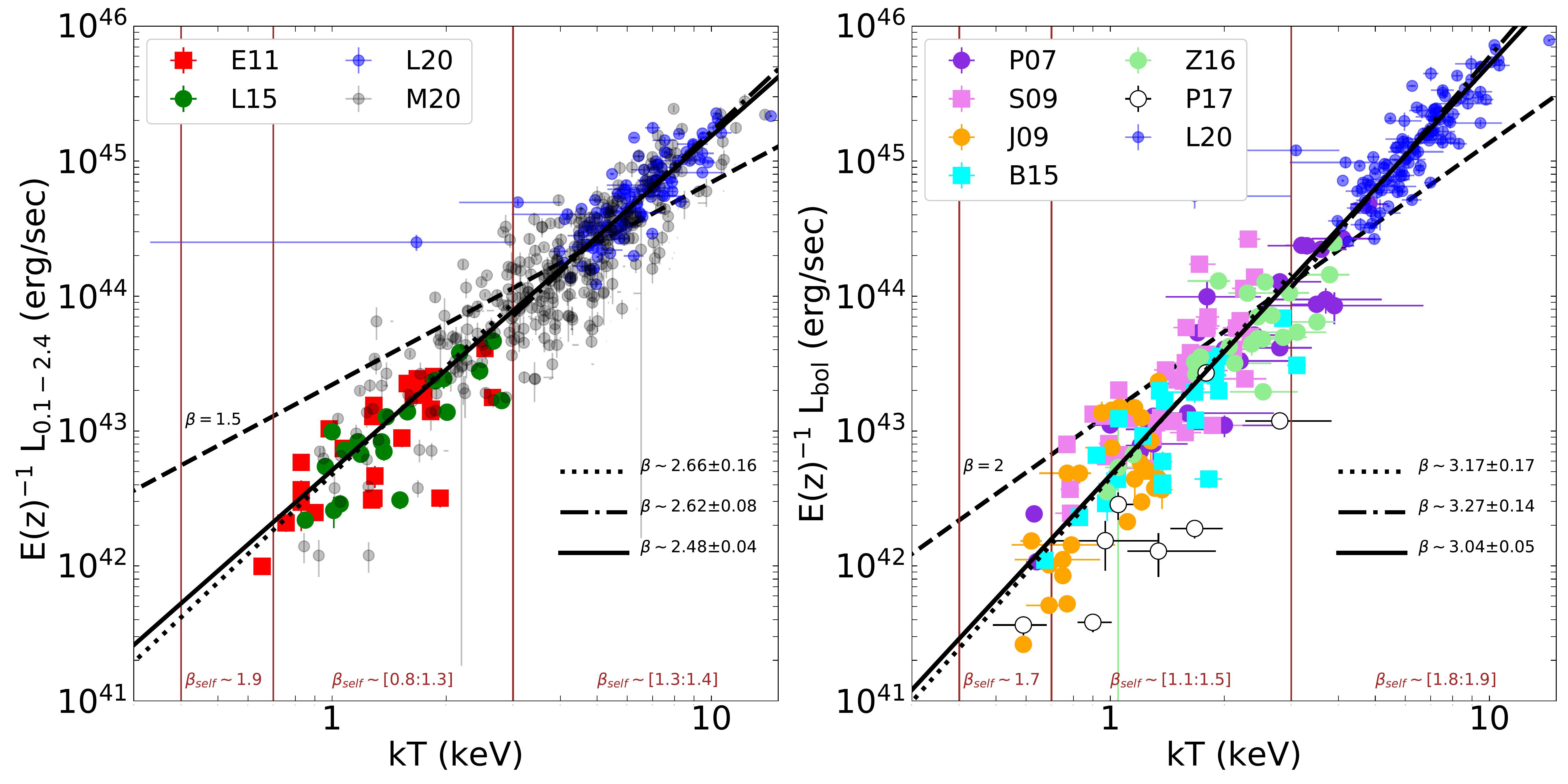}
\caption{{\it XMM-Newton} (circles) and {\it Chandra} (squares) measurements of the L$_{0.1-2.4}$--T$_{\rm x}$ ({\it \textbf{left panel}}) and L$_{\rm x,bol}$--T$_{\rm x}$ ({\it \textbf{right panel}}) relations for different samples of groups: \citet[E11]{Eckmiller},  \citet[L15]{LL15},  \citet[S09]{Sun2009}, \citet[J09]{Johnson2009}, \citet[B15]{Bharad2015}, \citet[Z16]{Zou2016}, \citet[P17]{pearson2017}. The groups measurements are compared with the ones from X-ray-selected (\citealt{Migkas2020}, M20) and SZ-selected (\citealt{LL20}, L20) cluster samples. We note that different studies used different atomic models (including APEC v1.3.1 which provide a significantly different modeling of the Fe-L line with respect to newer versions). The luminosities are all within R$_{500}$ while temperatures are obtained in different regions (see Table \ref{slsumary}). The {\it Chandra} measurements are converted to  {\it XMM-Newton}-like temperatures using the relations given in \citet{2015A&A...575A..30S}. Empty symbols are from  optically-selected samples. The lines represent the fitted relation for T$_{\rm x}<$3 keV systems (dotted), T$_{\rm x}>$3 keV systems (dashed-dotted), all systems (solid), and are compared with the case predicted by the self-similar scenario (dashed). The fits have been performed with LIRA (\citealt{LIRA}) assuming self-similar time evolution and, conservatively, without the scatter on the X variable, and are meant for visualization purposes only.  In brown we provide the expected values for the slope using the dependence of the emissivity tabulated in Table \ref{table:em} for different ranges of temperatures (showed as brown vertical lines).}
\label{fig:LT}
\end{figure} 

The L$_{\rm x}$--T$_{\rm x}$ relation involves two of the easiest quantities that can be derived using X-ray data. It was one of the first X-ray correlations to be studied and is still one of the most disputable scaling law between integrated observed properties. In fact, there have been conflicting reports in the literature about whether the relation for groups behaves as the one derived for massive clusters (i.e., whether groups are simply scaled-down versions of clusters or not). It has been clear for many years that the L$_{\rm x,bol}$--T$_{\rm x}$ relation for massive systems does not scale self-similarly (see, e.g., \citealt{Giodini2013} for a review about the relation for galaxy clusters), with slopes significantly higher than 2. Although pioneering studies of the relation for galaxy groups suggested considerably steeper slopes (i.e., slopes larger than 4;  \citealt{2000MNRAS.315..356H}, \citealt{helsdon00}, \citealt{xue00}), later investigations
found relations only slightly steeper than the ones for clusters (e.g., \citealt{osmond04}, \citealt{2009ApJ...690..879S}, \citealt{Eckmiller},  \citealt{Sun2012}, \citealt{LL15}). In Figure \ref{fig:LT}, we show a compilation of data for the L$_{\rm x}$--T$_{\rm x}$ relation taken from recent studies of galaxy groups observed with {\it XMM-Newton} and {\it Chandra}, and in Table \ref{slsumary} we list the best-fit slopes from these and other studies. The results show that indeed the slope obtained for poor systems (i.e., kT$<$3 keV) is consistent to the one derived for the more massive clusters (with hints of a slightly different normalization that cause a flattening when all the systems are fitted together). However, since the L$_{\rm x}$--T$_{\rm x}$ relation is expected to flatten in the low-mass regime (see Section \ref{selfsimilar}) these results clearly indicate a more significant contribution of the non-gravitational processes at the group scale. 
In fact, feedback processes (e.g, AGN heating) are expected to increase the entropy of the gas reducing its density (and hence the X-ray luminosity, steepening the relation). For massive systems the binding energy is so large that only the very central regions are affected, and the integrated properties of galaxy clusters remain essentially unchanged. Conversely, at the group scale the gas can be easily removed towards or beyond the virial radius modifying their global properties. The agreement between the L$_{\rm x}$--T$_{\rm x}$ relation of groups and clusters seems to stand also when the Malmquist bias (i.e., the preferential detection of intrinsically bright objects) is accounted for in galaxy groups studies as previously done for massive clusters. The Malmquist bias is expected to flatten the observed X-ray relations because only objects above a certain flux value are considered (either because one enforce an observational threshold or because faint systems are not detected). The correction needed to recover the underlying relation depends on the real intrinsic scatter with larger values requiring a larger correction (i.e., if the scatter increases in the low-mass regime then the magnitude of the flattening is larger than for galaxy clusters). While an attempt to correct this bias in cluster scaling relations has been provided in many different X-ray studies (e.g., \citealt{2002A&A...383..773I}, \citealt{2006ApJ...648..956S}, \citealt{Pacaud2007}, \citealt{2009ApJ...692.1060V}, \citealt{Pratt09},  \citealt{2011A&A...532A.133M}, \citealt{2017MNRAS.469.3738S}), there are only a few papers providing the correction in the galaxy group regime. For instance, \citet{LL15} analysing an X-ray flux-selected sample of local groups showed an increase  of the L$_{0.1-2.4}$--T$_{\rm x}$ relation slope after correcting for the Malmquist bias. A similar result was obtained by \citet{Bharad2015}, who estimated a correction for an archival sample of groups observed with {\it Chandra}. 
In contrast to these results, are the finding by \citet{Kettula2015} and \citet{Zou2016} who did not find any significant steepening after the bias correction. However, all the bias corrected relations obtained in the different studies show a great agreement (once they are converted into the same energy band). This agreement may suggest that the observational discrepancies arise from differences in the sample selection (which might cause the sample to be more or less biased). Once the biases are accounted for, then the results are not sensitive to the initial choices. \citet{Zou2016} showed that even once selection biases are taken into account the L$_{\rm x,bol}$--T$_{\rm x}$ relation at the group scale is consistent with the one for clusters. This finding confirms the stronger impact of the non-gravitational processes in the low-mass regime (otherwise a flattening should be observed).
Of course, these corrections work under the assumption that the X-ray selected samples are representative of the underlying population which might not be the case as suggested by, e.g., \citet{Rasmussen2006}, \citet{Anderson:2015}, \citet{Andreon2016}, and \citet{Ewan2017} who argued that the X-ray surveys miss a large fraction of galaxy systems. One possible reason for this incompleteness is related to the source detection algorithms mostly based on sliding cell detection methods. These algorithms work efficiently at finding point-like sources but has difficulties in detecting extended features, especially for nearby objects and for sources close to the detection limit (e.g., \citealt{2001A&A...370..689V}; see also \citealt{suhada12} for a performance comparison  between sliding cell and wavelet detection algorithms). \citet{2018A&A...619A.162X}, using a method optimized for the extended source detection, found a large number of new group candidates which are not included in any existing X-ray or Sunyaev-Zel'dovich (SZ) cluster catalogs. If  studies are restricted to groups that are a priori known to be X-ray bright and which properties may be quite different from those of optically-selected groups, as argued by \citet{Miniati2016}, then our view could be significantly biased.

\begin{table}[!t]
\caption{Overview of the most recent published scaling relations for galaxy groups based on {\it XMM-Newton} and {\it Chandra} data.}
\begin{tabular}{lcccccccc}
\toprule
\textbf{Relation} & \textbf{N} & \textbf{kT (keV)} & \textbf{z range} & \textbf{slope$_{self}$} & \textbf{slope$_{obs}$} & \textbf{slope$_{LIRA}$} & \textbf{reference} & \textbf{Note}  \\
\midrule
L--T$_{exc}$ & 26 & 0.6-3.0 & 0.012-0.049 & [0.9:1.1] & 2.25$\pm$0.21 & 3.27$\pm$0.26 & E11 & {\it a} ${\dagger}$ $\square$ $\upuparrows$ \\
L--T$_{exc}$ BC & 20 & 0.9-2.8 & 0.012-0.034 & [0.7:1.2] & 2.86$\pm$0.29  & - & L15 & {\it a} $\dagger$ $\square$ $\downdownarrows$\\
L--T$_{exc}$ & 20 & 0.9-2.8 & 0.012-0.034 & [0.7:1.2] & 2.05$\pm$0.32 & 2.90$\pm$0.36 & L15 & {\it a} $\dagger$ $\square$ $\downdownarrows$ \\
L$_{exc}$--T$_{exc}$ BC & 12 & 1.7-8.2 & 0.1-0.47 & [1.3:1.5] & 2.52$\pm$0.17 & - & K15 & {\it a} $\ddagger$ $\triangle$ $\downdownarrows$ \\
L$_{exc}$--T$_{exc}$ & 12 & 1.7-8.2 & 0.1-0.47 & [1.3:1.5] & 2.65$\pm$0.17 & 2.47$\pm$1.23 & K15 & {\it a} $\ddagger$ $\triangle$ $\downdownarrows$ \\
L--T$_{exc}$ BC & 26 & 0.6-3.6 & 0.012-0.049 & [1.2:1.6] & 3.20$\pm$0.26 & - & B15 & {\it c} $\dagger$ $\square$ $\upuparrows$ \\
L--T$_{exc}$ & 26 & 0.6-3.6 & 0.012-0.049 & [1.2:1.6] & 2.17$\pm$0.26 & 3.11$\pm$0.54 & B15 & {\it c} $\dagger$ $\square$ $\upuparrows$ \\
L--T BC & 23 & 1.0-3.9 & 0.03-0.147 & [0.8:1.3] & 2.79$\pm$0.33 & - & Z16 & {\it b} $\ddagger$ $\Diamond$ $\upuparrows$ \\
L--T BC & 23 & 1.0-3.9 & 0.03-0.147 & [1.2:1.6] & 3.29$\pm$0.33 & - & Z16 & {\it c} $\ddagger$ $\Diamond$ $\upuparrows$ \\
L--T & 23 & 1.0-3.9 & 0.03-0.147 & [1.2:1.6] & 3.28$\pm$0.33 & 2.92$\pm$0.25 & Z16 & {\it c} $\ddagger$ $\upuparrows$ \\
L$_{exc}$--T$_{exc}$ & 23 & 1.0-3.9 & 0.03-0.147 & [1.2:1.6] & 3.81$\pm$0.46 & 3.46$\pm$0.45 & Z16 & {\it c} $\ddagger$ $\Diamond$ $\upuparrows$ \\
\midrule
L--M$_{HE}$ & 26 & 0.6-3.0 &  0.012-0.049 & [0.4:0.8] & 1.34$\pm$0.18 & 1.47$\pm$0.43 & E11 & {\it a} $\dagger$ $\upuparrows$ \\
L--M$_{HE}$ BC & 20 & 0.9-2.8 & 0.012-0.034 & [0.2:0.7] & 1.66$\pm$0.22 & - & L15 & {\it a} $\dagger$ $\downdownarrows$ \\
L--M$_{HE}$ & 20 & 0.9-2.8 & 0.012-0.034 & [0.2:0.7] & 1.32$\pm$0.24 & 1.68$\pm$0.32 & L15 & {\it a} $\dagger$ $\downdownarrows$ \\
L$_{exc}$--M$_{WL}$ & 12 & 1.7-8.2 & 0.1-0.47 & [0.8:0.9] & 1.43$\pm$0.16 & 1.52$\pm$0.73 &  K15 & {\it a} $\ddagger$ $\downdownarrows$ \\
L--M$_{WL}$ BC & 105 & 0.6-6.0 & 0.054-1.033 & [0.4:0.8] & 1.07$\pm$0.37 & - & S20 & {\it b} $\ddagger$ $\downdownarrows$ \\
\midrule
M$_{HE}$--T$_{exc}$ & 43 & 0.7-2.7 & 0.012-0.122 & 1.5 & 1.67$\pm$0.15 & 1.75$\pm$0.14 & S09 & $\Diamond$ $\upuparrows$ \\
M$_{HE}$--T$_{exc}$ & 26 & 0.6-3.0 & 0.012-0.049 & 1.5 & 1.68$\pm$0.20 & 1.87$\pm$0.37 & E11 & $\square$ $\upuparrows$ \\
M$_{WL}$--T$_{exc}$ & 10 & 1.2-4.6 & 0.124-0.834 & 1.5 & 1.71$\pm$0.49 & 1.46$\pm$0.58 & K13 & $\circledcirc$ $\downdownarrows$ \\
M$_{HE}$--T$_{exc}$ & 20 & 0.9-2.8 & 0.012-0.034 & 1.5 & 1.65$\pm$0.07 & 1.61$\pm$0.10 & L15 & $\square$ $\downdownarrows$ \\
M$_{WL}$--T$_{exc}$ BC & 12 & 1.7-8.2 & 0.1-0.47 & 1.5 & 1.52$\pm$0.17 & - & K15 & $\triangle$ $\downdownarrows$ \\
M$_{WL}$--T$_{exc}$ & 12 & 1.7-8.2 & 0.1-0.47 & 1.5 & 1.68$\pm$0.17 & 1.22$\pm$0.82 & K15 & $\triangle$ $\downdownarrows$ \\
M$_{WL}$--T$_{300}$ & 76 & 0.6-6.0 & 0.044-1.002 & 1.5 & 1.33$\pm$0.75 & 1.14$\pm$0.32 & U20 & $\boxtimes$ $\downdownarrows$ \\
\midrule
M$_{HE}$--Y$_X$ & 43 & 0.7-2.7 & 0.012-0.122 & 0.6 & 0.56$\pm$0.03 & 0.71$\pm$0.24& S09 & $\upuparrows$\\
M$_{HE}$--Y$_X$ & 26 & 0.6-3.0 & 0.012-0.049 & 0.6 & 0.53$\pm$0.06 & 0.58$\pm$0.19 & E11 & $\upuparrows$ \\
M$_{HE}$--Y$_X$ & 20 & 0.9-2.8 & 0.012-0.034 & 0.6 & 0.60$\pm$0.03 & 0.58$\pm$0.04 & L15 & $\downdownarrows$\\
\midrule
M$_g$--M$_{HE}$ $^{\star}$ & 43 & 0.7-2.7 & 0.012-0.122 & 1 & 1.14$\pm$0.03 & 0.97$\pm$0.21 & S09 & $\upuparrows$\\
M$_g$--M$_{HE}$ & 26 & 0.6-3.0 & 0.012-0.049 & 1 & 1.38$\pm$0.18 & 1.22$\pm$0.44 & E11 & $\upuparrows$\\
M$_g$--M$_{HE}$ & 20 & 0.9-2.8 & 0.012-0.034 & 1 & 1.09$\pm$0.08 & 1.11$\pm$0.10 & L15 & $\downdownarrows$\\
M$_g$--M$_{WL}$ & 118 & 0.6-6.0 & 0.054-1.033 & 1 & 1.35$\pm$0.30 & - & S20 & $\downdownarrows$ \\
\midrule
K--T$_{exc}$ & 43 & 0.7-2.7 & 0.012-0.122 & 1 & 0.83$\pm$0.20 & - & S09 & $\Diamond$ $\upuparrows$ \\
\bottomrule
\end{tabular}
\label{slsumary}
\begin{tablenotes}
\setlength\labelsep{0pt}
\item {\scriptsize The subscripts {\it exc}, {\it 300}, {\it HE}, and {\it WL} indicate properties derived excluding the core, within R$<$300 kpc, under the assumption of hydrostatic equilibrium, and with weak-lensing analysis, respectively. BC indicates the relations corrected for selection effects. \\
The slope of the L$_{\rm x}$--T$_{\rm x}$ relation predicted by the self-similar scenario have been obtained as L$\propto$T$^{1.5+\gamma}$ where $\gamma$ is the slope of the X-ray emissivity in the considered energy band (e.g., soft or bolometric) and temperature range covered by the systems analyzed in each work (see Table \ref{table:em}). Since the X-ray emissivity strongly depends on the metallicity we provide the extreme values obtained with Z$_{\odot}$=0.3 and  Z$_{\odot}$=1.0. The slope of the L$_{\rm x}$--M relation is obtained similarly as L$_{\rm x}\propto$M$^{1+\gamma}$. \\
\noindent The values for slope$_{LIRA}$ have been obtained by fitting each dataset with LIRA (\citealt{LIRA}) assuming self-similar time evolution with scatter on both variables and with the following pivot values: 3 keV, 10$^{44}$ erg/sec,  2$\times$10$^{14}$ M$_{\odot}$, 10$^{13}$ M$_{\odot}$, 10$^{14}$ M$_{\odot}$ for T$_{\rm x}$, L$_{\rm x}$, M, M$_{\rm g}$, and Y$_{\rm X}$ respectively. Before fitting the L$_{\rm x}$--T$_{\rm x}$ relation, {\it Chandra} temperatures have been converted into {\it XMM-Newton}-like temperatures. \\
\noindent Note: {\it a}, {\it b}, and {\it c} refer to L$_{\rm x}$ obtained in the 0.1-2.4 keV, 0.5-2 keV, and bolometric band; $\dagger$ and $\ddagger$ indicate L$_{x}$ obtained with {\it ROSAT} or {\it XMM} data, while $\upuparrows$  and $\downdownarrows$ indicate if {\it Chandra} or {\it XMM} data have been used for the analysis.  $\square$, $\triangle$, and $\Diamond$, indicate that the core-excised region was not a fixed fraction of $R_{500}$, or fixed to 0.1$R_{500}$ or 0.15$R_{500}$, while $\circledcirc$ and $\boxtimes$ indicate the region 0.1-0.5R$_{500}$ and R$<$300 kpc, respectively. 
$^{\star}$ Relation derived fitting together the groups with a sample of  clusters. \\
\noindent References: \citet[S09]{Sun2009}, \citet[E11]{Eckmiller}, \citet[K13]{Kettula2013}, \citet[L15]{LL15}, \citet[B15]{Bharad2015}, \citet[K15]{Kettula2015},  \citet[Z16]{Zou2016},  \citet[U20]{Umetsu2020}, \citet[S20]{Sereno2020}.}
\end{tablenotes}
\end{table}

Beside the selection biases there are other issues complicating the comparison between different studies and between systems with different temperatures (masses). The first is the cross-calibration uncertainty between different instruments. For instance, \citet[see also \citealt{2010A&A...523A..22N}]{2015A&A...575A..30S} showed that the cluster temperatures derived with {\it XMM-Newton} are systematically lower than those obtained with {\it Chandra}. To complicate this issue is the temperature dependence of this difference. Fortunately, in the low temperature regime the differences are relatively small, a result that seems to hold also when including {\it Suzaku} data (e.g., \citealt{Kettula2013}). Although some caution is still needed, one can expect that calibration differences do not significantly affect the derived relations at the group scale. However, the impact of the calibrations needs to be taken into account when comparing the results obtained for sample of groups and sample of clusters. Another issue, pointed out by \citet{osmond04}, is related to the flattening of the fitted relation because the scatter in $\log{({\rm T_{x}})}$ will be asymmetric (assuming that the scatter in temperature is symmetric) with larger scatter towards low $\log{({\rm T_{x}})}$. Moreover, if the quality of the data are homogeneous across the sample, the statistical errors are expected to be larger in systems with low luminosities, which also tend to flatten the fitted relation. Finally, each study employ a different fitting algorithm (each with pros and cons) and treatment of the scatter and selection biases which impact the final results (see, e.g., \citealt{LL20}). In an attempt to remove this last uncertainty and provide comparable results we fit the published data with the same fitting method (i.e., using LIRA; \citealt{LIRA}) and assumptions (e.g., self-similar redshift evolution). The results are given in  Table \ref{slsumary}. 

The correlation between X-ray luminosity and gas temperature reflects  the fact that a deeper potential well (leading to a higher T$_{\rm x}$) generally contains more hot gas (leading to a higher L$_{\rm x}$). However, it has been shown that the gas fraction varies as function of the total mass with galaxy groups showing almost a factor of two lower gas fraction than galaxy clusters. Since the X-ray luminosity is proportional to the amount of gas in the IGrM and ICM, a change in the gas content in low-mass systems translates into a lower luminosity with the effect of steepening the L$_{\rm x}$--T$_{\rm x}$ relation. Anyway, the mass dependence of the gas fraction seems to vanish in the outer regions (e.g., \citealt{Sun2009}) implying that the low gas fraction observed in groups is mainly due to the low gas fraction of groups within $\sim$R$_{2500}$. This weak ability of the groups to retain the gas in the inner regions is probably a consequence of their shallow gravitational potential and thus of the increasing contribution of different non-gravitational processes.  These mechanisms are expected to provide an extra heating to the gas preventing the gas from falling toward the center, and by that, reducing gas density and X–ray emissivity in the cores. The effect is expected to play a significant role in poor systems leading to the steepening of the L$_{\rm x}$--T$_{\rm x}$ relation, as observed, and of the L$_{\rm x}$--$\sigma_{\rm v}$ relation which, however, is not currently supported by observations (see Section \ref{sec:lx-vdisp}). Supporting this scenario is the fact that when the core regions of galaxy clusters are ignored (i.e., by removing the regions where non-gravitational processes are expected to affect more the gas properties) the slope of the L$_{\rm x}$--T$_{\rm x}$ relation is more in agreement with the self-similar prediction. This is also in agreement with the suggestion by \citet{2011A&A...532A.133M} that, for low-temperature systems (i.e., kT$<$ 2.5 keV), AGN heating becomes more important than ICM cooling (which is the dominant mechanism in massive clusters). \citet{2007A&A...466..421C} suggested that intracluster magnetic fields can also affect more strongly the gas properties in the low-mass regime, resulting in an effective steepening of the scaling relations. 
In fact, as shown in \citet{2007A&A...466..421C}, the magnetic pressure tends to counterbalance part of the gravitational pull of the cluster preventing the gas from a further infalling. Thus, the presence of a magnetic field determines the final distribution of the gas density resulting in a less concentrated core (i.e., leading to a lower luminosity). The effect is mild for massive systems (due to their large gravitational potential) but is relevant in the group regime. The presence of the magnetic field is also expected to decrease the temperature because of the additional magnetic field energy term that needs to be included in the virial theorem. However, since galaxy groups and clusters are not isolated systems, the presence of an external pressure induced by the infalling gas from filaments, would tend to compensate the decrease of T$_{\rm x}$ caused by the magnetic field.

The non-gravitational heating implies the existence of an entropy floor (i.e., an excess of entropy with respect to the level referable to the gravity only) calling for some energetic mechanisms that can be summarized in three classes: preheating, local heating, and cooling (see the companion review by \citealt{2021Univ....7..142E} for detailed description of these mechanisms). Indeed, entropy in excess with respect to that achievable by pure gravitational collapse is observed in the inner regions of groups and poor clusters (e.g., \citealt{Mahdavi2005}, \citealt{Finoguenov2007}, \citealt{Johnson2009,Sun2009}, \citealt{2014MNRAS.438.2341P}). The excess is found to be radial and mass-dependent, being smaller for massive systems and extending to larger radii in low-mass objects. Moreover, \citet{Johnson2009} found that the excess is higher for groups with higher feedback (roughly estimated assuming that both the integrated feedback from SNe and AGNs scale with the stellar mass). Since entropy is expected to remain unchanged when neglecting non-gravitational processes, in the self-similar scenario it simply scales with the gas temperature. The finding by \citet{Sun2009} shows that the slope of the relation depends on the scaled radius at which the measurement is taken, and groups behave more
regularly in the outer regions (e.g., beyond R$_{2500}$) than in the core. This was already pointed-out by \citet{Ponman2003}. Thus, the slope of the K-T$_{\rm x}$ relation approaches the self-similar value at R$_{500}$, where there is no significant entropy excess above the entropy baseline (see the companion review by \citealt{2021Univ....7..142E}).  This is in agreement with the finding by \citet{Pratt2010} for a sample of galaxy clusters.

\begin{figure}[t]
\centering
\includegraphics[width=\textwidth]{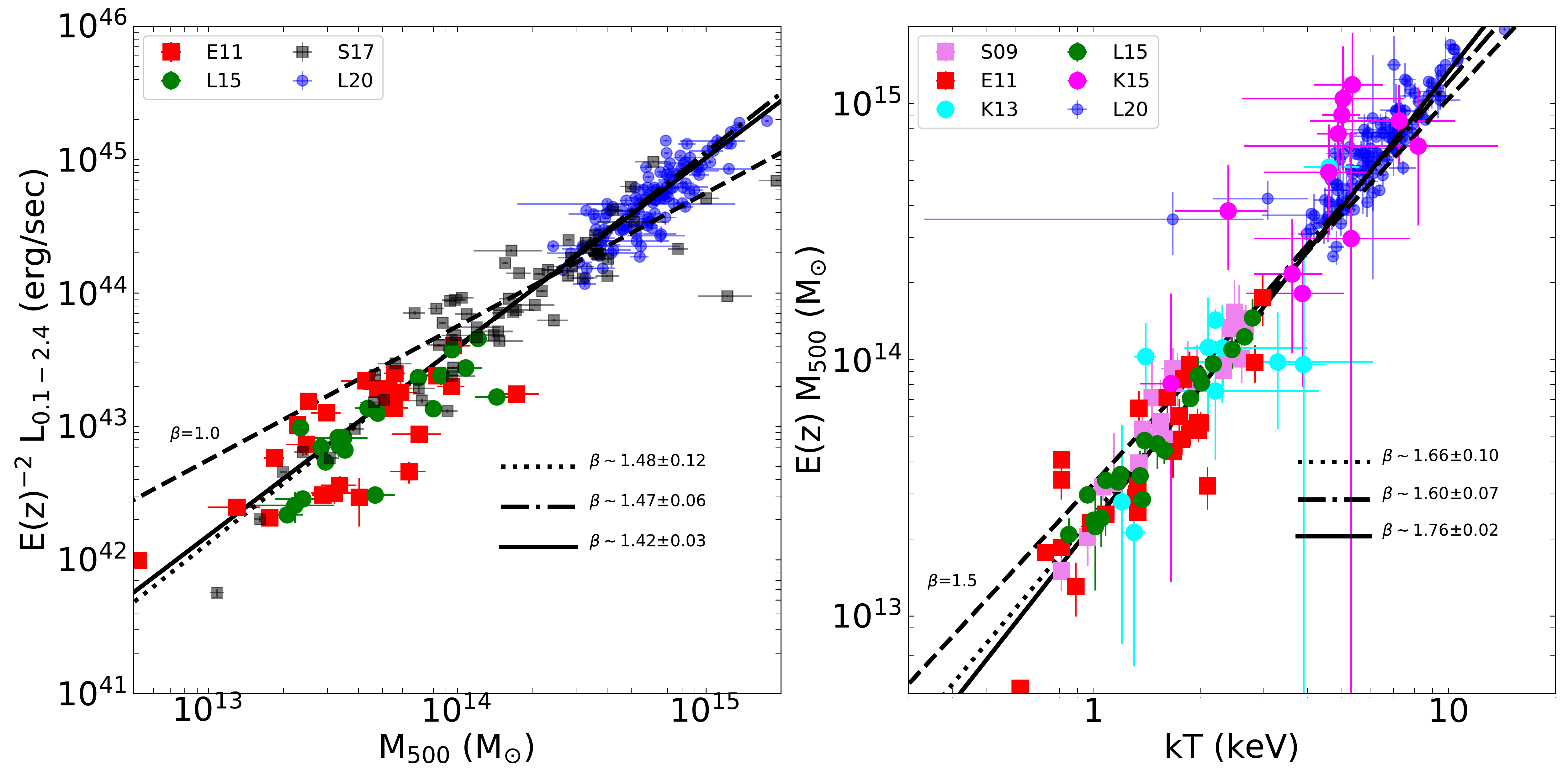}
\caption{{\it XMM-Newton} (circles) and {\it Chandra} (squares) measurements of the L$_{0.1-2.4}$--M ({\it \textbf{left panel}}) and M--T$_{\rm x}$ ({\it \textbf{right panel}}) relations for different samples of groups: \citet[E11]{Eckmiller}, \citet[L15]{LL15}, \citet[S09]{Sun2009} , \citet[K13]{Kettula2013}. The groups measurements are compared with the ones from an X-ray-selected (\citealt{2017MNRAS.469.3738S}, S17) and an SZ-selected (\citealt{LL20}, L20) cluster sample. Luminosities and masses are all within R$_{500}$ while temperatures are obtained in different regions (see Table \ref{slsumary}). Note that different studies used different methods to estimate the total masses.  The lines represent the fitted relation for M$<$10$^{14}$M$_{\odot}$ systems (dotted), M$>$10$^{14}$M$_{\odot}$ systems (dashed-dotted), all systems (solid), and are compared with the case predicted by the self-similar scenario (dashed). Because of the strong covariance between M and T$_{\rm x}$ we did not convert the {\it Chandra} temperatures to {\it XMM-Newton}-like temperatures.}
\label{fig:LM&MT}
\end{figure} 

Less studied than the L$_{\rm x}$--T$_{\rm x}$ relation, but of paramount importance for cosmological studies, is the relationship between X-ray luminosity and total mass (i.e., L$_{\rm x}$-M). This is particularly true for shallow X-ray surveys because it can be used to directly convert the easiest to derive observable (luminosity requires only source detection and redshift information) to the total mass. The calibration of this relation down to the low-mass regime will allow to break the degeneracy between $\Omega_{\rm m}$ and $\sigma_8$ (e.g., \citealt{Reiprich2002}). A large number of observations of galaxy clusters (e.g., see \citealt{2016MNRAS.456.4020M}, \citealt{2017MNRAS.471.1370S}, \citealt{2018MNRAS.473.3072M}, \citealt{Bulbul2019}, and \citealt{LL20} for recent studies; we refer to \citealt{Bohringer2012} and \citealt{Giodini2013} for older investigations) found that most of the values for the relation slope range from 1.4 to 1.9, steeper than the self-similar prediction of 4/3 suggesting that the luminosity is affected by non-gravitational processes. Unfortunately, the literature in the low-mass regime is still quite limited (in the {\it left} panel of Figure \ref{fig:LM&MT} we show a compilation of recent galaxy groups studies).  \citet{Eckmiller} found a slope of 1.34$\pm$0.18 and suggested that the single power-law modeling of the relation holds also for low-mass objects (see also the illustrative fit in Figure \ref{fig:LM&MT}).  However, given the considerations provided in the previous section, for the sample analyzed by \citet{Eckmiller} the luminosity should scale with the total mass to the power of [0.4:0.8] (see Table \ref{slsumary}) significantly lower than the finding by \citet{Eckmiller}. A similar slope was also obtained by \citet{LL15} but, after correcting for the selection biases, the corrected slope is steeper (i.e., 1.66$\pm$0.22) than the observed one (i.e., 1.32$\pm$0.24). making the deviations from the self-similar prediction even larger than what observed for galaxy clusters. This behavior can be explained by a gradual steepening of the true underlying L$_{\rm x}$--M relation towards the low-mass regime.  This implies that, as expected, also the luminosity of groups is heavily affected by non-gravitational processes. However, one should keep in mind that a mass-dependent bias in the total mass can also affect the shape of the L$_{\rm x}$--M relation. Because of the difficulties to distinguish the low temperature emitting gas of these systems from the Galactic foreground, the properties of galaxy groups can usually be observed out to a smaller radial extent than what is done for galaxy clusters. Thus, an estimate of the group masses at R$_{500}$ requires an extrapolation for most of the systems making the groups more prone to biases. Moreover, mass biases can also arise from the assumptions of hydrostatic equilibrium and spherical symmetry which are not valid for most of the systems. One way to overcome the problem is to use the weak-lensing masses which are expected to provide a less biased view of the true masses. Because of the difficulties to obtain shear maps for low-mass systems the first attempts have been performed via stacking analysis. \citet{2010ApJ...709...97L} stacked the weak lensing measurements of a sample of X-ray-selected galaxy groups and found an L$_{\rm x}$--M$_{200}$ relation in agreement with the finding of galaxy clusters. This result suggests that the L$_{\rm x}$--M$_{200}$ relation is well described by a single power-law down to the low-mass regime.  However, the lensing analysis of galaxy groups by \cite{Kettula2015} led to a shallower slope although the agreement in the low-mass regime of the two relations is fairly good, with significant tension appearing only at high masses (i.e., above a few 10$^{14}$ M$_{\odot}$). With the advent of multi-wavelength surveys, which uniformly scan large areas of sky, there has been significant progress in the weak-lensing analysis of large samples of galaxy groups. For instance, in the XXL  framework (see \citealt{Pierre2016}),  the L$_{\rm x}$--M$_{\rm WL}$ relation was investigated by \citet{Sereno2020} who found a quite good agreement with previous X-ray studies. The results suggest that the measured hydrostatic bias is consistent  with a small role of non-thermal pressure. However, due to the large uncertainties associated to the derived weak-lensing masses a large deviation from hydrostatic equilibrium cannot be completely excluded and further investigations with larger samples and higher quality data are required to make progress in the field. 

In contrast to the L$_{\rm x}$--T$_{\rm x}$ and L$_{\rm x}$--M relations, the M--T$_{\rm x}$ relation is expected to follow the same behavior for galaxy groups and clusters, under the assumption that the gas temperature reflects the depth of the underlying potential well. However, while the assumption is probably reasonable for many clusters (at least for the most relaxed ones) it may not be strictly true for groups where the gas has probably been significantly heated by non-gravitational processes.  For this reason, the expectation is that the scatter should increase in the low-mass regime where the global temperature is not insensitive to the details of the heating/cooling processes as in the high-mass regime. However, these processes are definitely more important in the inner regions with their effect fading at large distances from the center. Nonetheless, some simulations suggested that the gas removed by AGN activity in groups can affect the gas properties out to several Mpc (e.g., \citealt{2010MNRAS.402.1536S}), potentially affecting also cosmic shear measurements (e.g., \citealt{2011MNRAS.417.2020S}). Thus, it is important to define the region within which the characteristic cluster temperature is determined. This is not trivial because, for instance,  we have evidence that the central drop (i.e., the region in the center of relaxed clusters showing a significant temperature decline, probably caused by radiative cooling), typically present in relaxed clusters, does not scale uniformly with the mass (\citealt{2010A&A...513A..37H}). However, a common practice is to exclude the regions within 0.15R$_{500}$. 

There had been a lot of studies investigating the M--T$_{\rm x}$ relation before the {\it Chandra} and {\it XMM-Newton} era. Many of them (e.g., \citealt{2001A&A...368..749F}, \citealt{Sanderson2003}, and references therein) suggested that the low- and high-mass end of the relation is characterized by different slopes with the cross-over temperature between the two regimes at $\sim$3 keV. However, most of these studies could not constrain the gas properties (i.e., gas density and temperature gradients) at large radii making the estimated hydrostatic masses more prone to biases. Thanks to {\it Chandra} and {\it XMM-Newton}, the measurements could be extended to larger fraction of R$_{500}$ reducing the impact of the extrapolation. \citet{Sun2009} found a relation only slightly steeper than the prediction of the self-similar scenario. Both \citet{Eckmiller} and \citet{LL15} found that the slope for galaxy groups is consistent with the one of galaxy clusters but with a normalization 10-30\% lower. The net effect of this finding is a steepening of the relation when groups and clusters are fitted together (see, e.g., {\it right panel} of Figure \ref{fig:LM&MT}). Indeed, due to the limited field-of-view of {\it Chandra} and the high and variable {\it XMM-Newton} background level, X-ray measurements are still not tracing well the outer regions (despite the improvement with respect to previous missions, measurements extend out to R$_{500}$ only for a few systems). If the density profiles of groups are steepening at large radii then the masses could be underestimated explaining the lower normalization of the M--T$_{\rm x}$ relation. 
Another possible issue is that the samples analyzed in the above-mentioned papers are biased toward relaxed systems. Thus, if relaxed and disturbed systems do not share the same relation (e.g., because hydrostatic masses are more biased for disturbed systems) the relative fraction of relaxed/disturbed groups can impact the normalization (and possibly the slope) of the observed M--T$_{\rm x}$ relations. \citet{LL20} showed that this is probably not the case for massive systems, but a dedicated study in the group regime is still missing. Focusing on the slopes, the results of the most recent papers on galaxy groups are in agreement with the results for galaxy clusters for which most of the slope values range from 1.5 to 1.7 (see Table \ref{slsumary}). This agreement suggests a small impact of the non-gravitation processes to this relation. Again, before overinterpreting these results, one of the key questions is to asses if the level of mass bias in these systems is similar to the one of galaxy clusters. For instance, \citet[see also \citealt{Kettula2015}]{Kettula2013_MT} argued that the hydrostatic mass bias at 1 keV reaches a level of 30\%-50\%, higher than what usually observed for galaxy clusters (e.g., by calibrating the hydrostatic masses with other mass proxies, like weak-lensing or velocity dispersion). However, the sample consists of only 10 galaxy groups and the dynamical state of the systems is not discussed. A much larger sample was investigated by \citet{Umetsu2020} who found the relation to be consistent within statistical uncertainties with the self-similar expectations. However, the uncertainties of the individual weak-lensing mass measurements in the group regime are still quite large and tighter constraints are needed in the future to exclude deviations from self-similarity. An increasing mass bias in the low-mass regime would not be fully unexpected. In fact, in the unmagnetized case, the viscosity scales as T$_{\rm x}^{5/2}$ (\citealt{1962pfig.book.....S}) favoring the development of strong turbulences. Acting as additional pressure support against gravity, turbulent motions may increase the mass bias. Anyway, the IGrM is magnetized and the real magnitude of the turbulence is still unknown. 

Beside the shape of the scaling relations another important information is given by the scatter (i.e., the dispersion around the best-fit). Minimizing the scatter of the scaling relations is of paramount importance to obtain accurate constraints on cosmological parameters which are dominated by uncertainties in the mass--observable relations. Moreover, understanding the scatter in the relations is the key to pinpoint the physical processes at play in the group regime. However, the measurement errors for most of the groups are large, thus the intrinsic scatter is not well constrained, yet. Because of that,  \cite{Sun2009} and \cite{LL15} were not able to constrain the intrinsic scatter in their samples. \cite{Eckmiller} instead suggested that the scatter in galaxy groups (kT$<$3 keV) is much larger than the one derived for the HIFLUGCS sample (\citealt{Reiprich2002}).

A mass proxy which has been shown by simulations to bear a low scatter is the Y$_{\rm X}$ parameter (i.e., the product of the gas temperature and the gas mass; see \citealt{Kravtsov2006}). \citet{Sun2009} were the first to investigate the M--Y$_{\rm X}$ relation in the low-mass regime finding that a single power law model can fit very well both galaxy groups and clusters. This result was also confirmed by \citet{Eckmiller} and \citet{LL15}. Unfortunately, given the sample size and the relatively large measurements errors, the intrinsic scatter is not properly constrained. However, both \citet{Sun2009} and \citet{Eckmiller} suggested that the scatter of the M--Y$_{\rm X}$ relation is almost half of the M--T$_{\rm x}$ relation. The findings by \citet{Eckmiller} also suggest that the scatter for galaxy groups is significantly higher than for galaxy clusters. 

The self-similar model also predicts the X-ray scaling relations to be redshift-dependent (e.g., \citealt{Giodini2013} and references therein), reflecting the decrease with time of the mean density of the Universe. Non-gravitational processes are expected to affect the evolution of the X-ray scaling relations because of the increasing importance of such processes to the energy budget of galaxy systems as a function of redshift. Unfortunately, although groups are more common than clusters, because of their fainter and cooler nature it is more difficult to detect them over the background, especially at higher redshifts. Thus, due to the big challenges to detect large and representative samples of galaxy groups beyond the local Universe the literature on this subject is very limited. The few studies (e.g., \citealt{Jeltema2006}, \citealt{Pacaud2007}, \citealt{2010MNRAS.407.2543A}, \citealt{Umetsu2020}, \citealt{Sereno2020}) which have tried to address the evolution of the X-ray properties of galaxy groups did not find convincing evidence for such evolution. 
A characterization of the evolution of the scaling relations also on galaxy group scales is one of the goal of the next generation instruments (such as Athena; see Section \ref{future}).

\section{Optical scaling relations} 
\label{s:multi}

Due to the low X-ray flux at the group scale, there is high probability that X-ray selected samples are biased toward groups with rich IGrM. Moreover, since the luminosity strongly depends on the metallicity (see Section \ref{selfsimilar}), variations in the metal abundance between groups (possibly related to their feedback history) can significantly impact the selection function. Thus, it is advantageous to explore scaling relations between an X-ray property that can be measured relatively well in the low count regime (e.g., X-ray luminosity or gas temperature) and an optical property that can be used as a proxy for the group mass (e.g., velocity dispersion, optical band luminosity). 

\subsection{Velocity Dispersion}
\label{sec:veldisp}

The velocity dispersion ($\sigma_{\rm v}$) of galaxy groups (and indeed clusters) can be used to estimate dynamical masses via the application of the virial theorem.  Furthermore, the velocity of member galaxies complements X-ray information about the cluster morphology projected onto the sky.  For example, studying the luminosity--velocity dispersion (\Lsigma) relation provides an understanding of the dynamical properties of galaxy clusters and their impact on the scaling relations.

One of the most commonly used estimators of the velocity dispersion at the group regime is via the use of the {\em gapper} estimator from \cite{beers90}.  Of critical importance at the group scale, the {\em gapper} estimator is unbiased when using low numbers of member galaxies (down to $\sim$10 members, e.g., \citealt{ruel14}), and is robust against outliers.  The {\em gapper} velocity estimator ($\sigma_{\rm v}$) is given by
\begin{align}
    \sigma_{\rm v} &= \frac{\sqrt{\pi}}{N(N -1)}\sum^{N - 1}_{i=1} w_{i} g_{i},
\end{align}
where, for ordered velocity measurements, the gaps between each velocity pair are defined as $g_{i}=v_{i+1} - v_{i}$ (for $i=1,2,3...,N-1$), as well as Gaussian weights defined as $w_{i}=i(N - i)$.  

As stated above, one can study the \Lsigma~relation to understand dynamical properties and the impact on scaling relations.  In Equation~\ref{eq:LT}, it is shown that in the self-similar scenario the bolometric luminosity is expected to scale with the gas temperature as L$_{\rm x,bol} \propto$T$^{2}_{x}$.  Under the consideration that both the cluster/group hot gas and galaxies feel the same potential, assuming that they both have the same kinetic energy, the temperature can be converted to velocity dispersion using
\begin{align}
\label{eq:betaT}
    \beta =  \frac{\sigma^{2}_{\rm v} \mu {\rm m_{p}}}{\rm k_{B} {\rm T_{x}}} \approx 1,
\end{align}
where the parameter $\beta$ is the ratio of the specific energy in galaxies to the specific energy in the hot gas.
Using Equation~\ref{eq:betaT} and the self-similar scaling of L$_{\rm x,bol}$ and T$_{\rm x}$ above, the self-similar scaling of velocity dispersion and X-ray properties can be given by
\begin{align}
\label{eq:Lsigma}
    {\rm L_{x,bol}} \propto \sigma^{4}_{\rm v},
\end{align}
\begin{align}
\label{eq:Tsigma}
    {\rm T_{x}} \propto \sigma^{2}_{\rm v}. 
\end{align}

However, because of the behavior of the X-ray emissivity in the low-temperature regime, the dependence of the luminosity on the temperature is more complicated (see discussion in $\S$\ref{selfsimilar}) and can be approximated as L$_{\rm x}\propto$T$_{\rm x}^{1.5+\gamma}$, where $\gamma$ is the slope of the X-ray emissivity in the considered energy band (e.g., soft or bolometric) and  temperature range (see Table \ref{table:em}).  Using this $\gamma$ dependent relation, it follows that 
\begin{align}
\label{eq:lsigma_corr}
L_{\rm x}\propto\sigma_{\rm v}^{3+2\gamma}. 
\end{align}
For temperatures lower than 3 keV, the value of $\gamma$ is negative (unless very cool systems are considered), implying that the expected \Lsigma~relation for galaxy groups is shallower than what is predicted for galaxy clusters (e.g., Equation \ref{eq:Lsigma}).

\subsection{The luminosity-velocity dispersion relation}
\label{sec:lx-vdisp}

\begin{figure}[!t]
\begin{center}
\includegraphics[width=0.75\textwidth]{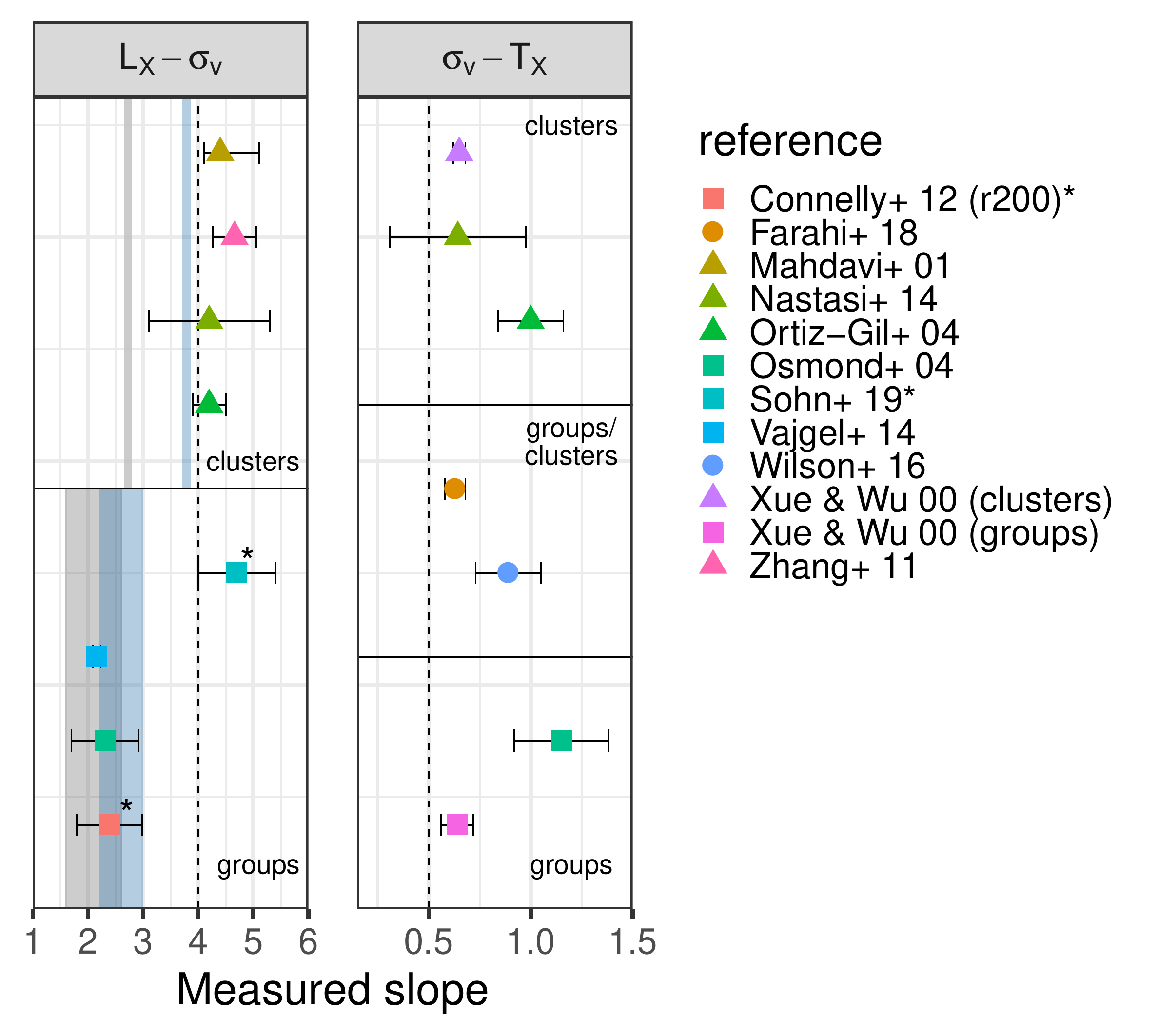}
\end{center}
\caption{Compilation of the measured slopes of the luminosity-velocity dispersion (\Lsigma, {\it \textbf{left panel}}) and velocity dispersion-temperature (\sigmaT, {\it \textbf{right panel}}) relation in the literature.  In each case, the dashed vertical line highlights the usual self-similar expectation on the slope for each relation (Equation~\ref{eq:Lsigma}, {\it \textbf{left panel}}, and Equation ~\ref{eq:Tsigma}, {\it \textbf{right panel}}).  The horizontal lines represent a dividing line between the mass scales used for the comparison in each relation.  As shown in Sect.~\ref{sec:veldisp} when the \Lsigma~scaling can be given as L$_{\rm x}\propto\sigma_{\rm v}^{3+2\gamma}$ (with $\gamma$ dependant on the energy band and emissivity).  The blue shaded region represents the self-similar expectation when considering bolometric luminosities for clusters (assuming T$_{\rm x}$=3.0-10.0 keV, L$_{\rm x}\propto\sigma_{\rm v}^{[3.7:3.9]}$) and groups (assuming T$_{\rm x}$=0.7-3.0 keV, L$_{\rm x}\propto\sigma_{\rm v}^{[2.2:3.0]}$).  The grey shaded region represents the self-similar expectation when considering 0.1--2.4 keV luminosities for clusters (assuming T$_{\rm x}$=3.0-10.0 keV, L$_{\rm x}\propto\sigma_{\rm v}^{[2.7:2.8]}$) and groups (assuming T$_{\rm x}$=0.7-3.0 keV, L$_{\rm x}\propto\sigma_{\rm v}^{[1.6:2.6]}$).  Unless otherwise stated, references consider (bolometric) luminosities and temperatures derived within an estimate of R$_{500}$. Please note that the \cite{sohn19} relation is based upon an analysis using 7 bins of using the full sample of 74 systems. * refers to references using 0.1-2.4 keV band luminosities.} 
\label{fig:slopes_comp}
\end{figure}

At the cluster scale, generally, it has been found that the observed luminosity--velocity dispersion (\Lsigma) relation follows, or is slightly steeper than, the expectation of Equation \ref{eq:Lsigma} (e.g., \citealt{Quintana1982,Mulchaey1998,mahdavi01,ortizgil04,zhang11,nastasi14}).  Furthermore, at the cluster scale, studies of the \Lsigma~relation now utilise samples of clusters numbering in the high hundreds (e.g., \citealt{kirkpatrick21}, using 755 clusters to investigate the \Lsigma~ relation). Studies of the \Lsigma~relation at the group scale attempt to compare the form of the relation at the high-mass regime in order to investigate differences at these two mass scales (e.g., to probe the effect of AGN feedback processes at high and low masses).  Early studies comparing the slope of the relation between the two mass regimes provided a mixed picture, with studies finding groups have a flatter (e.g.,  \citealt{helsdon00,xue00,osmond04}) or consistent (e.g., \citealt{Ponman1996}, \citealt{Mulchaey1998}, \citealt{mahdavi01}) relation than their high-mass counterparts.  Figure~\ref{fig:slopes_comp} ({\it left-panel}) provides a (non-comprehensive) compilation of the slope of the \Lsigma~relation from various studies in the literature.  The solid horizontal line represents the dividing line between studies using clusters (top half) and groups (bottom half).  The L$_{\rm x,bol} \propto \sigma^{4}_{\rm v}$ expectation is given by the vertical dashed line.  While there appears to be a clear division between the slopes for groups and clusters, many of the group scale studies compare to the usual L$_{\rm x,bol} \propto \sigma^{4}_{\rm v}$ expectation at the cluster scale.  As shown in Equation~\ref{eq:lsigma_corr}, the scaling can be given by L$_{\rm x}\propto\sigma_{\rm v}^{3+2\gamma}$, with $\gamma$ dependent on the X-ray emissivity and energy band used.  If we assume a group temperature range of 0.7--3.0 keV, then given the range of emissivities in Table~\ref{table:em}, the bolometric scaling in the group regime becomes L$_{\rm x,bol}\propto\sigma_{\rm v}^{[2.2:3.0]}$.  This range is highlighted by the blue shaded region in Figure~\ref{fig:slopes_comp} ({\it left-panel}) at the group scale.  For comparison, using these same arguments, the bolometric scaling for clusters (assuming 3.0-10.0 keV) becomes L$_{\rm x,bol}\propto\sigma_{\rm v}^{[3.7:3.9]}$ (again highlighted by the blue shaded region in Figure~\ref{fig:slopes_comp}, appropriate to the cluster scale).  Considering the above, studies investigating the group scale relation can indeed be considered consistent with the self-similar expectation (e.g., \citealt{osmond04}). 
While this is the case, many authors note caveats when studying groups, which are discussed below.

Many early studies were based upon ensemble collections of groups that can lead to biases in the derived scaling relations.  In recent years, studies of the \Lsigma~relation have utilised groups selected over contiguous survey regions.  One such study was performed by \cite{connelly12}.  Groups were selected from regions of the Canadian Network for Observational Cosmology Field Galaxy Redshift Survey 2 (CNOC2,  \citealt{carlberg99}) that were covered by {\it XMM-Newton} and {\it Chandra} observations, totalling 0.2 and 0.3 deg$^{2}$ contiguous areas of two fields of the CNOC2 survey.  Using X-ray selected groups with high quality redshift information, they find a slope of the \Lsigma~of 2.40$^{+0.58}_{-0.60}$ (including groups with lower quality redshift information yields a slope of 1.35$^{+0.42}_{-0.47}$).  While initial inspection of the value of the slopes would imply the slope is shallower than the self-similar expectation (as noted in \citealt{connelly12}), we note that the luminosities are 
reconstructed in the 0.1--2.4 keV band (from the flux in the 0.5-2 keV band, and by correcting for extension and K-correction as described in \citealt{finoguenov07cosmos}).
 Assuming the scaling follows L$_{\rm x} \propto \sigma^{3+2\gamma}_{\rm v}$, then for luminosities in the 0.1--2.4 keV band, the scaling can be given by L$_{\rm x} \propto \sigma^{[2.0:2.7]}_{\rm v}$ (depending on the metallicity of the groups).  This expectation is shown in Figure~\ref{fig:slopes_comp} ({\it left-panel}), highlighted by the grey shaded region at the group regime.  The slope determined by \cite{connelly12} is coincident with this scaling.  Therefore, if the energy band is considered, the \cite{connelly12} relation is consistent with the self-similar expectation.  Hence, it can be assumed that the groups studied in \cite{connelly12} are consistent with the cluster scale (assuming clusters follow the self-similar expectation).  Another study utilising contiguous fields is presented in \cite{sohn19}, using groups selected from the 2 deg$^{2}$ Cosmic Evolution Survey (COSMOS, \citealt{scoville07}).  This study constructs a catalog of galaxy groups based upon those identified in \cite{george11}, using {\it XMM-Newton} and {\it Chandra} observations of the COSMOS field, reaching an X-ray flux limit of $\sim$10$^{-15}$ erg s$^{-1}$ cm$^{-2}$.  \cite{sohn19} use galaxy redshift information from a wide variety of surveys in the literature and associate them with the X-ray detected groups, compiling a final sample of 146 groups with at least three spectroscopic members.  Based upon a cleaned sample of 74 groups, \cite{sohn19} showed that the relation for individual groups appears to follow a shallower relation than clusters (consistent with that found by previous studies, e.g., \citealt{mahdavi01}).  However, they note that this trend may be affected by a small number of groups that appear to have anomalously low velocity dispersions (at $\sigma_{\rm v} \lesssim$125 km s$^{-1}$) for their measured X-ray luminosity (discussed further in Sect.~\ref{sec:low_vdisp}).  To overcome this, \cite{sohn19} estimated the median velocity dispersion for 7 bins of groups created from the 74 groups in their sample, finding a slope of L$_{\rm x} \propto \sigma^{4.7\pm0.7}_{\rm v}$.  This study considers luminosities in the 0.1--2.4 keV band, therefore, as discussed above, the measured slope is in fact steeper than the self-similar expectation.

Although contiguous regions have been used to study the \Lsigma~relation, an extremely small number of studies have attempted to correct for X-ray selection biases.  One study that attempts to do so is presented in \cite{vajgel14}, using 14 groups with at least 5 galaxy members selected from the 9 deg$^{2}$ X-Bo{\"o}tes survey (\citealt{murray05}).  They find that the group scale relation is consistent with the self similar expectation.  In order to test the effects of Malmquist bias on the observed relations, \cite{vajgel14} determined the limiting X-ray luminosity in two survey volumes ($z$=0.20 and $z$=0.35).  The resulting relations are consistent with the sample relation, with the authors concluding the sample may not be dominated by Malmquist bias effects.  However, due to the associated large error on each relation, making this conclusion is challenging and requires the construction of larger samples.  Another use of contiguous surveys, particularly those covered by multiple wavelengths, is the possibility to compare the relations derived using groups selected via multiple selection methods (e.g., X-ray, optical).  Furthermore, the use of optically selected groups allows one to estimate the form of scaling relations independent of the usual X-ray selection biases (e.g.,  \citealt{Andreon2016}).  The study by \cite{connelly12}, as detailed above, also constructed a sample of 38 optically (spectroscopically) selected groups.  Using this optically selected sample, they derive a slope of the \Lsigma~relation of 1.78$^{+0.60}_{-0.54}$.  Given the large uncertainties on the measured slopes, the comparison of the X-ray and optically selected samples is somewhat limited (note that the comparison is not effected by the energy band used, as discussed above, since they are consistent between the X-ray and optically selected samples).

\subsection{The velocity dispersion-temperature relation}
\label{sec:tx-vdisp}

Velocity dispersion and gas temperature are two independent probes of the depth of the cluster potential well, estimated by using baryons as tracers. Therefore, this relation can provide useful information about the effect of non-gravitational processes,  which are responsible for the deviation from thermal equilibrium of the IGrM and ICM. Hence, it is useful to compare group and cluster relations to investigate the differences between these mass scales. Figure~\ref{fig:slopes_comp} ({\it right-panel}) shows a compilation (again, a non-comprehensive picture) of the slope of the $\sigma_{\rm v}$-T$_{\rm x}$ relation from studies in the literature.  The horizontal lines represent the division between studies using groups (bottom section), clusters (top section) and those using systems which straddle the group/cluster regime (middle section).  Based upon Equation~\ref{eq:betaT}, it is expected that the velocity dispersion of the galaxies should scale with the square-root of the temperature of the gas, $\sigma_{\rm v} \propto$T$_{\rm x}^{1/2}$.  In the context of clusters, various studies have found that the \sigmaT~relation has a slope steeper than the self-similar expectation (e.g., \citealt{ortizgil04,wilson16}), with others finding a steeper slope but with errors too large to confirm a deviation (e.g., \citealt{nastasi14}). Unfortunately, the study of the \sigmaT~relation for groups is somewhat limited in the literature.  An early investigation presented in \cite{osmond04} showed evidence for steepening of the relation at the group scale (where they find a slope of 1.15$\pm$0.23).  However, they caution that there is both large uncertainties on the measured X-ray temperatures and a large amount of scatter observed in the relation, which could be the cause of tension with previous studies attempting to investigate any steepening of the relation for groups.  Although \cite{osmond04} found evidence for a steepening of the relation, they remarked that a comparison cluster based relation passes through the centre of the group relation data, and represents adequately the cluster based relation.  However, recent studies of the relation, especially at the group scale, become scarce.  One recent study of the $\sigma_{\rm v}$-T$_{\rm x}$ is presented in \cite{wilson16}, making use of groups/clusters detected serendipitously in the {\em XMM} Cluster Survey (\citealt{romer99}).  Using 19 groups/clusters having redshifts z$<$0.5, spanning the temperature range 1.0 $\lta$ T$_{\rm x}$ $\lta$ 5.5 keV (with 50\% of clusters having a temperature $<$3 keV), the \sigmaT~relation is found to have a slope of 0.89$\pm$0.16.  While again steeper than the self-similar expectation, the result is somewhat shallower (although not significant) than that presented in \cite{osmond04}.  Finally, the last relation considered is that given in \cite{farahi18XXL}, which investigated the \sigmaT~relation for a sample of X-ray selected clusters detected in the XXL survey (\citealt{Pierre2016}).  Clusters were selected from the 25 deg$^{2}$ XXL-N region, with spectroscopic data compiled from a range of surveys (see \citealt{adami18}, for full details of the spectroscopic coverage).  Using a sample of 132 clusters (the majority of which have T$_{x}<$3 keV), \cite{farahi18XXL} found a relation of the form $\sigma_{\rm v} \propto$T$_{\rm x}^{0.63\pm0.05}$.  Note that the relation is fitted using an ensemble maximum likelihood method, with fitted slope in tension with the self-similar expectation.  Since both velocity dispersion and X-ray temperature scales with total mass, one can combine the information to determine a useful mass calibration (e.g., \citealt{farahi18XXL}).  However, consideration must be given to velocity anisotropies during mass modelling using velocity information, which can vary for loose, compact and viralised groups (\citealt{mamon03}).  However, corrections based upon halo concentration have been developed (e.g., \citealt{mamon13}).  
The study of the $\sigma_{\rm v}$-T$_{\rm x}$ relation can also be probed down to the galaxy scale.  \cite{goulding16} used galaxies from the volume-limited MASSIVE survey (\citealt{ma14}), to study the relation between galaxy kinematics ($\sigma_{\rm e}$) and X-ray temperature.  \cite{goulding16} found a relation of the form T$_{\rm x} \propto \sigma_{\rm e}^{1.3-1.8}$ (note the inverse of the relation as discussed above), noted as being marginally flatter than the self-similar expectation.

As discussed in Section~\ref{sec:low_vdisp}, AGN feedback and its effects on the ICM could result in deviations from self-similarity, in particular, the steepening observed above in the \Tsigma.  While an observational consensus on the magnitude of the deviation from self-similarity at the group scale compared to the cluster scale has yet to be reached, simulations have indeed shown a mass dependence (e.g., \citealt{lebrun17}, \citealt{farahi18}, \citealt{Truong:2018}).  The deviations discussed for the \Lsigma~and \sigmaT~are thought to arise due to the effects of AGN feedback on the ICM (as shown in simulations, e.g., \citealt{chowdhury20}), which has little effect on the galaxy velocity dispersions.  Furthermore, \cite{goulding16} measured a median value of $\beta$=0.6 for their galaxy sample, suggesting the galaxies have undergone, or still in the process of, additional heating due to, e.g., AGN feedback, as discussed above.

\subsection{Low velocity dispersion groups}
\label{sec:low_vdisp}

One observation made by various authors studying the \Lsigma~relation, is the presence of low velocity dispersion groups (appearing at $\sigma_{\rm v} \lesssim$200 km s$^{-1}$) that have a high X-ray luminosity in comparison to their $\sigma_{\rm v}$ (conversely, it can be stated that these groups have a low $\sigma_{\rm v}$ for their L$_{\rm x}$).  These low velocity outliers have been noted in various studies in the literature (e.g., \citealt{mahdavi01,helsdon05}), attributed as the cause of the flattening of the \Lsigma~relation at the group scale (e.g., \citealt{vajgel14}, \citealt{sohn19}).  While it has been shown in Section~\ref{sec:lx-vdisp} that the group scale relations may be consistent with self-similar predictions when accounting for the differing emissivity, the presence of these outliers are extreme cases.  A physical interpretation of the low $\sigma_{\rm v}$ outliers is therefore currently lacking.  Furthermore, the presence of these outliers remains somewhat of a mystery if one considers the effects of AGN feedback on the intragroup medium.  During an AGN outburst, gas will be removed from the group potential, hence lowering the group overall X-ray luminosity.  With the group velocity dispersion unaffected by this process, the expectation would be that the group should have a lower L$_{\rm x}$ for a given $\sigma_{\rm v}$, contrary to this outlier population.  It could therefore be argued that it is in fact the velocity dispersion that have been underestimated for these groups.  Potential explanations for the presence of these low $\sigma_{\rm v}$ outliers were given in \cite{helsdon05}.  It is postulated the cause could be: (i) through dynamical friction, energy is transferred from a large orbiting body to the sea of dark matter particles through which it moves; (ii) due to tidal interactions, the orbital energy may be converted into internal energy of the galaxies; and (iii) the orbital motion  happens in the plane of the sky, therefore contributing little to the line of sight velocity dispersion.  
While a current physical interpretation is lacking, the presence of low velocity dispersion outliers could be due to X-ray selections effects (e.g., Eddington and Malmquist biases).  Extreme outliers (e.g., \citealt{sohn19}) may not be attributed to selection, however various relations involving L$_{x}$ when using X-ray selected samples characteristically show a flattening when not account for selection (e.g., \citealt{mantz10,giles17}). In fact, the preferential selection of higher luminosity groups for a given velocity dispersion (i.e., Malmquist bias), leads to the presence of ``moderate'' outliers.  \cite{Andreon2016} argues that an unbiased sample of clusters can be obtained when selecting clusters from optical properties and therefore able to probe the full range of scatter and the true form of the relation.  As stated, \cite{connelly12} have used an optically selected sample of clusters to investigate the form of the \Lsigma~relation, however, they find constancy in both the form and scatter of the X-ray and optically selected group samples (due to the large errors on the scaling parameters).  This sample only covered an area of 0.5 deg$^{2}$, therefore the comparison of optically and X-ray selected samples over overlapping contiguous fields requires further attention to truly probe the differences in selection.       

\begin{figure}
\begin{center}
\includegraphics[trim=0.0cm 0 0 0, clip, width=0.7\linewidth]{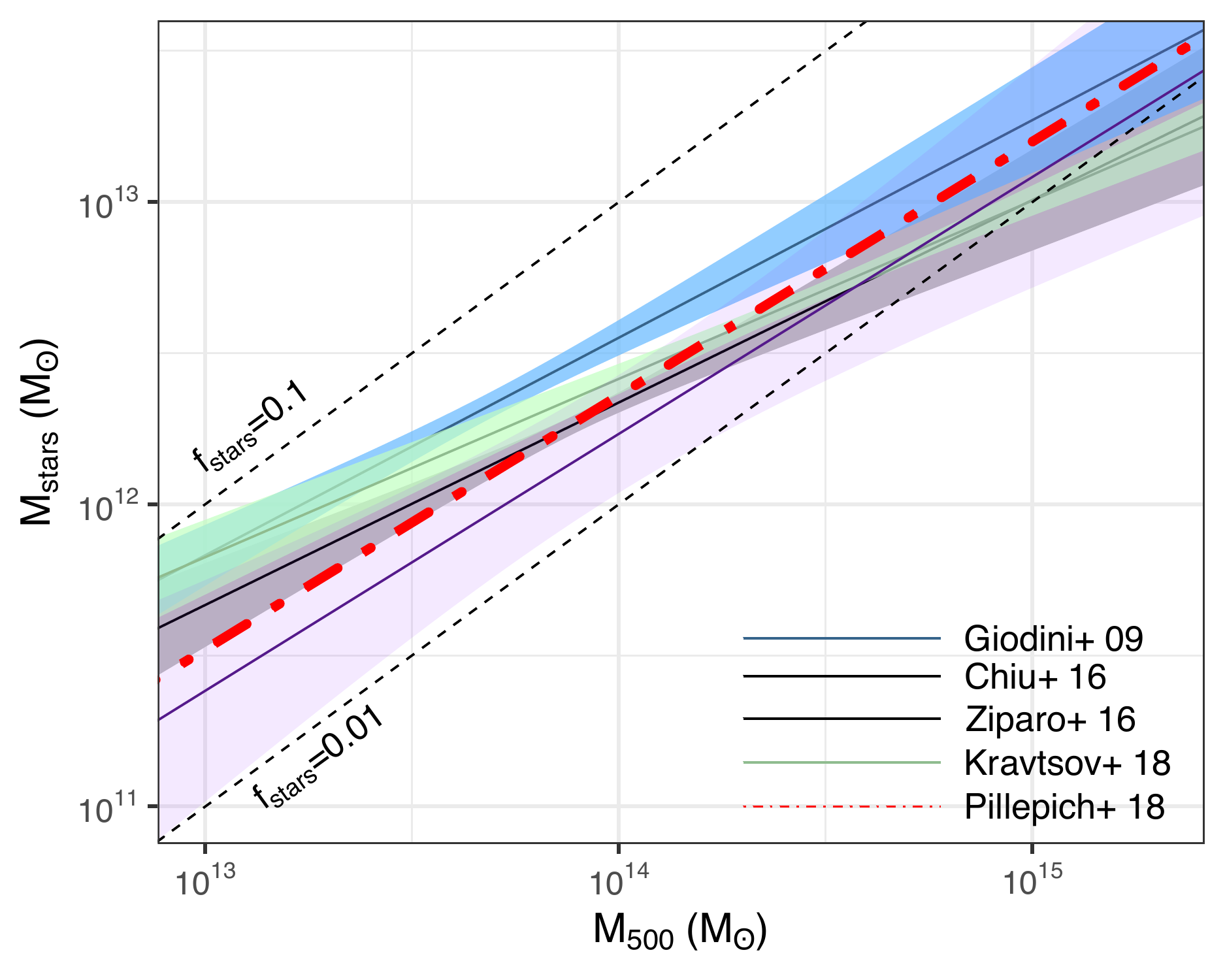}
\end{center}
\caption{The M$_{\rm stars}$--M relation of various studies in the literature.  Two lines of stellar mass fraction are highlighted by the dashed lines.  Please note that the \cite{ziparo16} relation is derived from the conversion of L$_{\rm K}$ to M$_{\rm stars}$ assuming a constant mass-to-light ratio of 0.73 (as used in \citealt{ziparo16}).} 
\label{fig:mstar_m}
\end{figure}

\subsection{Stellar gas content of galaxy groups}
\label{sec:stellargas}
The gas mass fraction of clusters can be used as a probe of cosmology (e.g., \citealt{Allen08}, \citealt{Ettori2009}, \citealt{mantz14}, \citealt{2017MNRAS.471.1370S}).  However, as mentioned in the previous sections, it has been shown that the fraction decreases as a function of total mass.  Interestingly, the opposite is true for the stellar mass fraction (f$_{\rm stars}$=M$_{\rm stars}$/M$_{\rm tot}$), with an increasing stellar mass fraction  as a function of decreasing total mass (e.g., \citealt{lin03}, \citealt{gonzalez2007}, \citealt{giodini09}, \citealt{2010ApJ...717..379B}, \citealt{2011A&A...535A..78Z}, \citealt{Leauthaud2012}, \citealt{2013A&A...555A..66L}, \citealt{2018MNRAS.478.3072C}, \citealt{2019ApJ...878...72D}).  To investigate this trend, much effort has been afforded to the study of the gas mass and stellar mass content in groups and clusters (e.g., to determine star formation rates).  One such observation is that the stellar mass has a correlation with the halo mass with a slope $<$1 (see discussion below).  This has the implication that at the group scale, star formation is more efficient.  One early study that specifically used groups to constrain the form of the stellar mass--halo mass relation (M$_{\rm stars}$--M) and the group f$_{\rm stars}$ is that of \cite{giodini09}.  An X-ray selected sample of groups was constructed from the COSMOS survey, in which X-ray extended sources were detected based upon a wavelet detection routine (\citealt{vikhlinin98b}).  Mean photometric redshifts were assigned to each candidate and checked against available spectroscopic redshifts from $z$COSMOS (\citealt{lilly07}).  After quality checks, a final sample of 91 groups were used to constrain the form of the M$_{\rm stars}$--M relation.  Masses were estimated based upon a stacked weak lensing analysis (\citealt{2010ApJ...709...97L}) and the construction of a \LMtwo~relation, from which the catalog masses were estimated (note that M$_{500}$ masses were used in the final analysis, estimated from the M$_{200}$ assuming an NFW profile -\citealt{Navarro1996}, \citealt{Navarro1997}-, and constant concentration, $c=5$).  Within R$_{500}$ the \MstarM~relation was found to follow a form of M$_{\rm stars}\propto$M$^{0.81\pm0.11}$ and a stellar mass fraction of the form f$_{\rm stars}\propto$M$^{-0.26\pm0.09}$ (extending this to higher masses with the inclusion of clusters, the form follows a relation of f$_{\rm stars}\propto$M$^{-0.37\pm0.04}$).  More recent studies have utilised increased area X-ray surveys.  The {\em XMM} {\em Blanco} Cosmology Survey (XMM-BCS, \citealt{suhada12}) covers 12 deg$^{2}$ of the sky with {\em XMM-Newton}, and was used by \cite{chiu16} to study the form and evolution of the \MstarM~relation using 46 groups/clusters within a mass and redshift range of ($2\lesssim {\rm M} \lesssim 25)\times10^{13}~{\rm M_{\odot}}$ and 0.1 $\lesssim z \lesssim$ 1.02, respectively.  The \MstarM~relation is fitted including an evolutionary redshift term, with parameters estimated by evaluating a likelihood based upon observing a cluster with observed properties (L$_{\rm x}$ and M$_{\rm stars}$) given a mass, redshift, \LM~relation (mass calibration) and the \MstarM~relation.  The likelihood is weighted by the mass function, with full details given in \cite{liu15}.  The fitted relation has the form M$_{\rm stars}\propto$M$^{0.69\pm0.15}(1+z)^{-0.04\pm0.47}$, again consistent with previous results showing a shallower than unity slope of the relation.  Further, these results indicate little evolution in the stellar content with stellar mass fraction staying constant out to $z\simeq$1.  In Figure~\ref{fig:mstar_m}, we plot the \MstarM~relation for various results obtained in the literature (namely, \citealt{giodini09,chiu16,ziparo16,kravtsov18,pillepich18}). We note that no attempt has been made to correct for the differences in mass calibration used in the various studies.  For reference, two constant stellar mass fractions are given by the black dashed lines.  As discussed above, the relations for \cite{giodini09} and \cite{chiu16} are derived at the group scale, whereas \cite{ziparo16} straddles the high-mass groups/low-mass cluster regime (see below) and \cite{kravtsov18} used primarily high-mass clusters (note that the relation plotted here includes clusters from \citealt{gonzalez13}, as detailed in \citealt{kravtsov18}).  All the relations have a slope less than unity, and show the trend of decreasing stellar mass fraction from the group to cluster regime.  The observed relations are also consistent with that found in simulations.  Results obtained from the IllustrisTNG simulations show the same trend in stellar mass (shown by the red dot-dashed line in Figure~\ref{fig:mstar_m}, taken from \citealt{pillepich18}).   

Due to the difficulty of measuring the stellar masses of groups directly, requiring deep observations, it is beneficial to use a proxy for the stellar mass.  One such proxy as a tracer of the stellar mass is the K-band luminosity (L$_{\rm K}$), as shown in various studies (e.g., \citealt{lin03,muzzin07}).  The use of L$_{\rm K}$ as a stellar mass proxy was investigated in \cite{ziparo16} using a sample of 20 groups/clusters selected from the {\em XXL} survey.  The clusters were selected from the overlap of the {\em XXL} and CFHTLS, using clusters with an individual weak lensing mass estimate.  \cite{ziparo16} found a relation of the form L$_{\rm K}\propto$M$^{0.85^{+0.35}_{-0.27}}_{\rm WL}$, which while shallower than unity, a slope of 1 cannot be ruled out.  Furthermore, when combined with a sample of high-mass clusters from LoCuSS, \cite{ziparo16} measured a slope of 1.05$^{+0.16}_{-0.14}$.  The relation derived for the {\em XXL} sample is shown in Figure~\ref{fig:mstar_m} (purple line, with the shaded light purple region highlighting the 1$\sigma$ uncertainty), which is derived from the L$_{\rm K}$--M$_{\rm WL}$ relation assuming a constant mass-to-light ratio of 0.73 (as adopted in \citealt{ziparo16}).

\section{The role of SMBHs: observed scaling relations and predictions via HD simulations}
\label{s:coev}
As introduced in Section \ref{s:intro}, the evolution of the IGrM filling galaxy groups can not be merely understood in isolation as giant self-similar gaseous spheres. Particularly in the last decade, a wide range of evidences have accumulated showing that the SMBHs at the center of each galaxy group are tightly co-evolving with the hot X-ray halo. Such co-evolution works in both directions: the hot halo acts as an active atmosphere and reservoir of gas which recurrently feeds the central SMBH (\citealt{Gaspari:2013_cca,Prasad:2015,Voit:2017,Temi:2018,Tremblay:2018,Gaspari:2018,Rose:2019,Storchi-Bergmann:2019}). In turn, the SMBH re-ejects back large amount of mass and energy (in particular via jets and outflows; e.g., \citealt{Tombesi:2013,Sadowski:2017,Fiore:2017}), thus re-heating and re-shaping the IGrM via bubbles, shocks, and turbulence up to the group outskirts (\citealt{McNamara:2012,Fabian:2012,Gitti:2012,Brighenti:2015,Gaspari:2015_xspec,Liu:2019,Yang:2019,Wittor:2020,Voit:2020}). 
While the small-scale AGN self-regulation thermodynamics/kinematics is respectively tackled in the companion \citet{2021Univ....7..142E} and \citet{Gastaldello:2021}~reviews, here we focus on the macro-scale integrated (X-ray) IGrM properties and group scaling relations, which complete and complement Sections \ref{s:sss} and \ref{s:multi}.
Furthermore, we compare with high-resolution and hydrodynamical (HD) simulations, in particular to discuss what the X-ray scaling relations can constrain and tell us in terms of the baryonic physics shaping the IGrM.

\begin{figure}[!t]
     \subfigure{\hspace{-2.9cm}
     \includegraphics[width=1.34\textwidth]{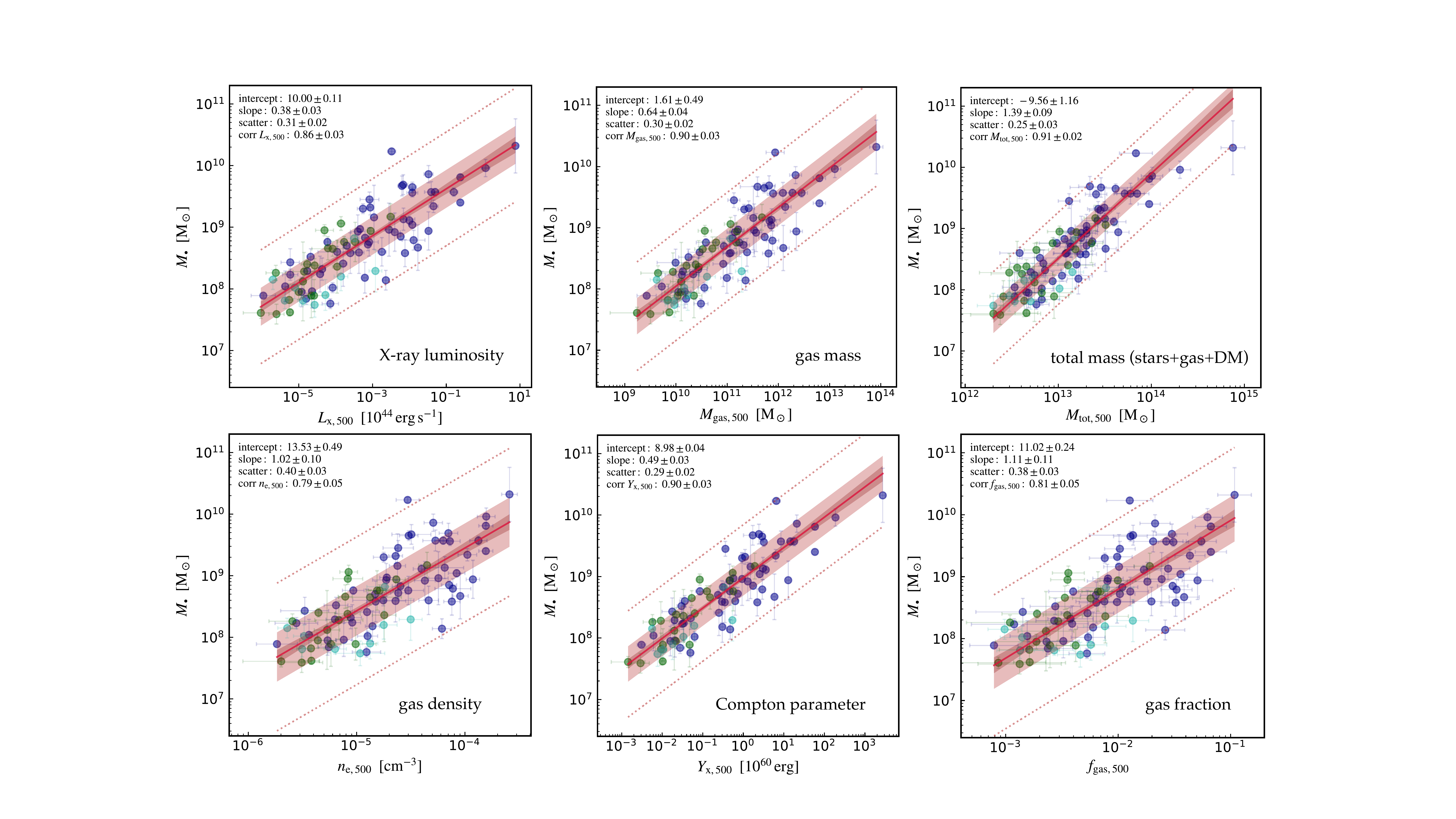}}
     \vspace{-0.8cm}
     \caption{Scaling relations between the central (dynamical/direct) SMBH mass and key macro X-ray halo properties --- adapted from \citet{Gaspari:2019}. 
     {\it \textbf{Top-left}} to {\it \textbf{bottom-right panels}}: gas X-ray luminosity (in the 0.3\,-\,7.0 keV band), 
     gas mass, total mass (dominated by the dark matter component), gas electron density, Compton parameter, and gas fraction. 
     The extraction radius is within the macro-scale $r < \r500$.
     The 85 circles include the observed direct/dynamical SMBH mass with the X-ray halo detected in the host galaxy groups (and a handful of clusters). The circle colors depict the morphological type of the central galaxy: elliptical/blue, lenticular/green, spiral/cyan. 
     The employed Bayesian analysis accounts for the observed errors, providing a statistically robust estimate of the intercept and slope of the linear fitting (top-left inset). The solid red line with dark band shows the 16-84 percentile interval of the fit, while the 1-$\sigma$ intrinsic scatter is shown via the wider light red band (dotted lines are the 3-$\sigma$ levels). 
     }
     \label{f:comp}
\end{figure}

Figure~\ref{f:comp} shows several key macro X-ray halo scaling relations, which are usually employed in cosmological studies (see Section \ref{s:sss}), but now plotted against the SMBH mass \Mbh, which is also an integrated property. 
These SMBH masses are retrieved only via robust direct measurements, i.e.~resolving the stellar or gas kinematics within the SMBH influence region (e.g., via HST). The current largest sample correlated with the available X-ray hot gas properties is presented by \citet{Gaspari:2019}, which includes central galaxies and satellites, with morphological types such as ellipticals (blue circles), lenticulars (green), and a few spirals (cyan). The 85 systems span a range of M$_{500} \sim 3\times10^{12}-3\times10^{14}$  M$_{\odot}$, with the majority of systems in the group regime (T$_{\rm x}\sim 1$ keV) and a few in the poor or cluster tails.
The companion \cite{2021Univ....7..142E}~review shows that the \Mbh\ correlation with $\tx$ is significantly tighter than the classical optical scalings, such as the Magorrian relation (e.g., \citealt{Kormendy:2013,Saglia:2016}), with intrinsic scatter down to 0.2\,dex, in particular within the circumgalactic and core region. Here, in Figure~\ref{f:comp} we show the other key X-ray properties integrated up to \R500, namely the plasma X-ray luminosity (in the 0.3\,-\,7.0 keV band), gas mass, total mass (gas plus stars plus dark matter), gas density, Compton parameter, and gas fraction. All the fits parameter -- including the intercept, slope, scatter, and correlation coefficient -- are shown in the top-left inset.
The related Bayesian analysis (\citealt{Gaspari:2019})
shows the 1-$\sigma$ intrinsic scatter as light red bands with the dotted lines enveloping the rare 3-$\sigma$ loci. As indicated by all correlation coefficients, even the macro-scale IGrM (several 100 kpc to Mpc scale) is tightly linked to the central \Mbh.
The tighter correlations are those involving the gas mass/luminosity, X-ray Compton parameter Y$_{\rm x}$, and total mass, while the loosest one is that with the gas density. 
It is interesting to note that using the core radius (or smaller) as extraction radius (not shown) leads to similar results, except that the total mass scatter increases by 0.1 dex, with the gas properties emerging as dominant drivers (in particular M$_{\rm gas}$ and f$_{\rm gas}$).
In other words, we suggest to use the \R500\ scaling to probe the total mass, while smaller extraction radii to probe gas mass (and related properties).

\begin{figure}[ht]
     \centering
     \subfigure{\hspace{-0.2cm}
     \includegraphics[width=0.7\textwidth,angle=270]{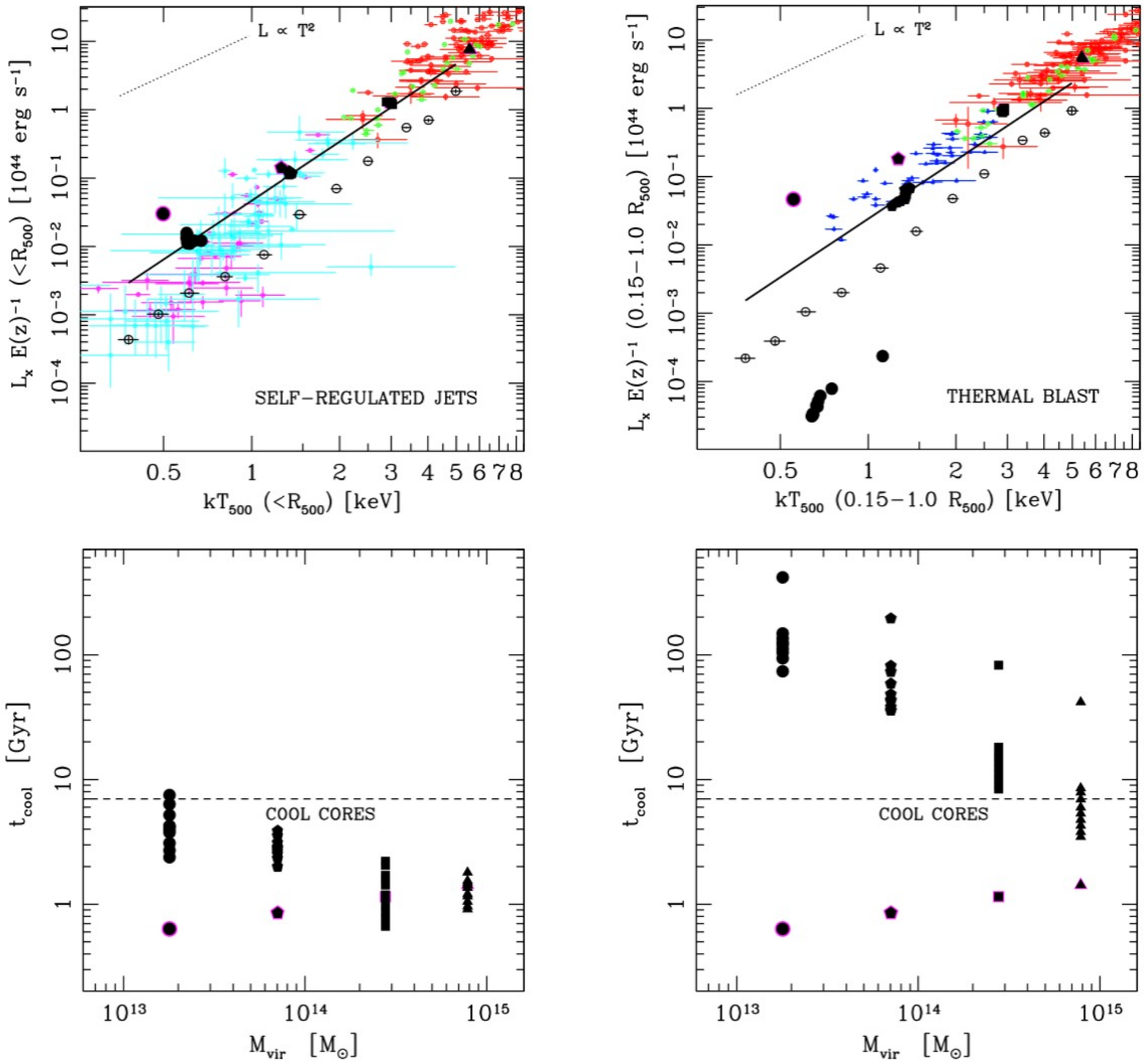}}
     \vspace{-1cm}
     \caption{Effects of different baryonic models in shaping the evolution of the hot halos, in particular the X-ray luminosity -- temperature relation ({\it \textbf{top panels}}) and cool-coreness via the central cooling time ({\it \textbf{bottom panels}}); adapted from \citet{Anderson:2015} and \citet{Gaspari:2014_scalings}.
     The colored individual objects in the top panels are from a wide range of observational works
     (\citealt{2000MNRAS.315..356H,Mulchaey:2003,osmond04}, \citealt{Sun2009,Pratt09}, \citealt{Maughan:2012}; luminosities are extracted mostly in the 0.5\,-\,2 keV band). The empty circles and solid line show the raw \citet{Anderson:2015} stacking analysis and the unbiased fit, respectively.
     The filled black points show evolutionary tracks in large-scale HD simulations (\citealt{Gaspari:2014_scalings}) implementing self-regulated AGN jets ({\it \textbf{left}})  or strong thermal blast feedback ({\it \textbf{right}}), preserving or evacuating the surrounding diffuse gaseous halo, respectively. The initial state is marked with magenta contour.
     Evacuation and overheating becomes particularly dramatic in low-mass, less-bound groups.
     }
     \label{f:models}
\end{figure}

The X-ray correlations shown in Figure~\ref{f:comp} are important to probe models of galaxy group evolution. 
A key debated topic in current extragalactic astrophysics is which mode of accretion feeds internally the IGrM and eventually the central SMBH. 
In hot accretion modes (usually Bondi or ADAF; e.g., \citealt{Bondi:1952,Narayan:2011}), the larger the thermal entropy of the gas, the stronger the feeding is stifled, as the inflowing gas has to overcome the hot-halo thermal pressure, increasing toward the center. This would induce negative correlations with the IGrM properties, which are ruled out by the slopes shown in Figure~\ref{f:comp}. Conversely, cold-mode accretion -- typically in chaotic form due to the turbulent IGrM condensation generating randomly colliding clouds (e.g., \citealt{Gaspari:2013_cca,Prasad:2015,Voit:2018,Olivares:2019}) -- would produce major positive and tight correlations with the gas mass and X-ray luminosity (e.g., the cooling rate is $\propto$ L$_{\rm x}$). Therefore, X-ray correlations favour chaotic cold accretion (CCA) over hot mode accretion.
Hierarchical mergers (of both SMBHs and galaxies) are another channel to potentially grow such correlations. However, cosmological simulations (\citealt{Bassini:2019,Truong:2021}) show this to be effective only at the high-mass end. Moreover, Figure~\ref{f:comp} shows that all the mass scalings are either sub- or super-linear, far off from any simple self-similar expectation. In other words, a positive baseline due to hierarchical assembly is present, but gas feeding (dominated by CCA, in terms of mass) substantially shapes the slope and scatter of such \Mbh\ correlations over the long-term evolution.
Overall, observed scaling relations of macro X-ray halo properties (shown in Sections \ref{s:sss} and \ref{s:multi}) cannot be thought as disjointed from scaling relations of micro properties (e.g., \Mbh), since both systems are tightly co-evolving and intertwined through the several billion years evolution and over 10 orders of magnitude in spatial scale (cf.~the diagram in \citealt{Gaspari:2020} linking the micro, meso, and macro scales). As striking as it appears, such scaling relations allow us to convert back and forth between vastly different scales, depending on the availability of either the micro (Section \ref{s:coev}) or macro (Section \ref{s:sss}) properties for each detected galaxy group.

The X-ray scaling relations presented in Section \ref{s:sss} can be also leveraged to test feedback models in large-scale simulations or to calibrate semi-analytic models of group evolution, thus giving us hints on the dominant baryonic processes in the IGrM (e.g., \citealt{Puchwein:2008}, \citealt{McCarthy:2010,Kravtsov:2012,Tremmel:2017}). 
Figure~\ref{f:models} shows the key impact of archetypal feedback models on the evolution of the diffuse hot atmospheres (\citealt{Gaspari:2014_scalings}). The filled black points indicate the Gyr evolution of the hot halos IGrM and ICM as it suffers recurrent injections of either anisotropic mechanical energy via jets (left column) or a strong impulsive thermal quasar-like blast (right column). Evidently, the latter model has a dramatic impact on the main \LT~relation (even when the core is excised), producing a catastrophic evacuation of gas that lowers luminosities by 3 orders of magnitude, especially toward lower-mass group regime ($\tx<1$ keV). Such quasar-like models are inconsistent with the observed X-ray scaling relations, in particular those probing the very low-mass regime via stacking analysis (e.g., \citealt{Anderson:2015} shown via empty circles and solid line fit). 
Conversely, a tight self-regulation (e.g., achieved via CCA feeding) and a flickering injection via gentle AGN jets can preserve the hot halo throughout the several 100 outburst cycles. The bottom panels show indeed that the initial cool core (magenta contour) can be preserved even in less bound halos, such as poor galaxy groups, without becoming overheated above half of the Hubble time. Such overheating is instead catastrophic for an impulsive AGN blast injection, transforming all hot halos into perennial non-cool-core systems, which is ruled out by observations finding groups to have almost universally a low central $t_{\rm cool}$ (\citealt{Sun2009,Babyk:2018_entropy}).
Such self-regulated, gentle SMBH feedback has thus become a staple for subgrid models of cosmological simulations which can reproduce other tight scaling relations without any major break at the group scale, such as the \MY~or \MT~computed over \R500 (e.g., \citealt{Planelles:2017,Truong:2018,Weinberger:2018}).
For comparisons with further cosmological simulations, we refer the interested reader to the companion reviews \citet{universe7070209}~and \citet{2021Univ....7..142E}~(section 5).

\section{Galaxy groups with the next generation instruments}
\label{future}
Over the next decade, dedicated survey instruments will increase the number of known groups and clusters out to high redshift, constraining the scenario for their formation and evolution. Examples include {\it eROSITA} in X-rays, {\it Vera Rubin Observatory} and {\it Euclid} in the Optical/Infrared, and several “Stage 3” ground-based mm-wave observatories. The SZ-effect surveys, in particular, will break new ground by providing robustly selected, large catalogs of clusters at $z > 1.5$, as well as the first informative absolute mass calibration from CMB-cluster lensing. 
All future observatories list the baryonic mass and energy distribution on groups' scales resolved up to redshift $\sim2$ and beyond, when they first appeared as collapsed X-ray bright structures, as one of their main scientific goals.

Currently, a big step forward in the collection and characterization of low-mass systems is expected from the ongoing observations of the all X-ray sky with the {\it extended ROentgen Survey with an Imaging Telescope Array} ({\it eROSITA}\footnote{http: //www.mpe.mpg.de/erosita/}, \citealt{2021A&A...647A...1P}).
eROSITA is operating in the X-ray energy band (0.2-10 keV) at L2 orbit on-board the ‘Spectrum-Roentgen-Gamma’ (SRG) satellite. {\it eROSITA} has a spatial resolution comparable to the {\it XMM–Newton} one, a similar effective area at low energies, but a wider field of view, while it will be 20–30 times more sensitive than the {\it ROSAT} sky survey in the soft band and will provide the first all sky imaging survey in the hard band. Optimizing galaxy group and cluster detection has been one of the most important tasks during the mission preparation (e.g., \citealt{2018A&A...617A..92C}, and \citealt{kaefer20} in particular for a detection and characterization through ICM outskirts that reduces possible biases due the peaked X-ray emission associated to cool cores). During its 4-yr-long all-sky survey, with an average exposure of 2.5 ks (whereas the average exposure in the ecliptic plane region is $\sim$1.6 ks), eROSITA is planned to deliver a sample of about 3 million active galactic nuclei (AGNs) and about 125,000 galaxy systems (mostly groups) detected with more than 50 photons and M$_{\rm 500c} > 10^{13} {\rm M}_{\odot}/h$ up to redshift $\sim1$ (median: $z\sim0.3$) (\citealt{Merloni:2012,2012MNRAS.422...44P,Borm2014,2018MNRAS.480..987Z,Pillepich:2018}). 
Almost all groups (and clusters) detected with {\it eROSITA} will lack sufficient X-ray photons to accurately constrain temperature and mass profiles (\citealt{Borm2014}). Thus, cosmological studies using group and cluster of galaxies to be detected with {\it eROSITA}, will rely heavily on a detailed understanding of the scaling relations where systematic effects would have to be factored in to ensure that the cosmological applications of these relations are not hampered. Hence, a thorough investigation of these systems, to understand the interplay between the development of the hot IGrM and feedback processes, becomes highly important, not only for cosmology but also to understand complex baryonic physics.
Moreover, to reach the planned goals of $1 \sigma$ errors of 1\%, 1\%, 7\%, and 25\% on $\sigma_8$, $\Omega_{\rm m}$, $w_{\rm 0}$, and $w_{\rm a}$, respectively, the critical passages will be: {\it i}) a better knowledge (by a factor of $\sim4$) of the parameters describing the L$_{\rm x}$--M relation to improve the constraints on $\sigma_8$ and $\Omega_{\rm m}$, and {\it ii}) a lower mass threshold to enlarge the analyzed sample to reduce the statistical uncertainties in DE sector.

The physics of IGrM and ICM will be the main scientific driver for the exposures with the {\it Advanced Telescope for High ENergy Astrophysics} ({\it Athena}\footnote{https://www.the-athena-x-ray-observatory.eu/}), the X-ray observatory mission selected by ESA as the second L(large)-class mission (due for launch in early 2030s) within its Cosmic Vision programme to address the Hot and Energetic Universe scientific theme. 
Among the main scientific goals, {\it Athena} will have the capabilities to find evolved groups of galaxies with M$_{\rm 500c} > 5\times 10^{13}$M$_{\odot}$ and hot gaseous atmospheres at $z>2$. 
For about ten of those, a global gas temperature estimate is expected to be measurable (\citealt{Pointecouteau:2013}). {\it Athena} will determine the magnitude of the injection of non-gravitational energy into the IGrM and ICM as a function of cosmic epoch by measuring the structural properties (e.g., the entropy profiles) out to R$_{500}$, and their evolution up to $z\sim2$, for a sample of galaxy groups and clusters, improving significantly the constraints, presently unknown, on the evolution of the scaling relations between bulk properties of the hot gas (\citealt{Ettori:2013_athena,Pointecouteau:2013}).
In local systems, {\it Athena} will be also in condition to determine the occurrence and impact of AGN feedback phenomena by searching for ripples in surface brightness in the cores of a statistical sample of objects. Using temperature-sensitive line ratios, {\it Athena}'s observations will trace how much gas is at each temperature in the cores of these systems, providing a complete description of the gas heating-cooling balance (\citealt{Croston:2013}) and transport processes such as turbulence and diffusion (\citealt{Cucchetti:2018,Roncarelli:2018,Mernier:2020}).

Presently, concepts funded for study by NASA for consideration in the 2020 Astrophysics Decadal Survey, {\it Lynx}\footnote{https://www.lynxobservatory.com/} (as high-energy flagship mission) and {\it AXIS}\footnote{http://axis.astro.umd.edu} (as probe-class mission) are proposing to investigate with sub-arcsecond resolution over a FoV of 400-500 arcmin$^2$ the X-ray sky, improving this capability of a factor $\sim100$ with respect to {\it Chandra} ACIS-I. 
Their predicted low background level and capability to resolve embedded and background AGN will allow to track group and cluster emission at very low surface brightness values. For example, {\it AXIS} is expected to reach a flux limit of $\sim 1 \times 10^{-16}$ erg/s/cm$^2$ (0.5--2 keV) over the 50 deg$^2$ of the proposed Wide Survey (e.g., \citealt{Marchesi:2020}), providing the detection of thousands of groups and clusters, and evidences of merging and effects of feedback resolved even at high-$z$.
With a larger collection of instruments, {\it Lynx} will be also able to resolve the thermodynamic and kinematic structure of systems at $z\approx 2$, as well as determine the role of feedback from AGN and stars.

Complementary data will be provided from the ongoing (and planned) SZ surveys.
{\it SPT-3G}\footnote{https://pole.uchicago.edu/} will extend the work of {\it SPT-SZ} by covering a nearly identical area of 2,500 deg$^2$ but with noise levels about 12, 7, and 20 times lower at 95, 150, and 220 GHz, respectively. 
This will enhance the sensitivity, allowing to reduce the mass limit and extending the redshift coverage with respect to {\it SPT-SZ}. About 5000 clusters with M$_{\rm 500c} \gta 10^{14}$M$_{\odot}$ at a signal-to-noise $> 4.5$ (corresponding to a 97\% purity threshold) are expected by the completion of the survey (2023; \citealt{Benson:2014}). 
The next-generation ground-based cosmic microwave background experiment {\it CMB-S4}\footnote{https://cmb-s4.org/}, with a planned beginning of science operations in 2029, will build catalogs more than an order of magnitude larger than current ones, lowering the mass limit M$_{\rm 500c}$ to 6-8 $\times 10^{13}$ M$_{\odot}$ at $z>0.3$ and being especially adept at finding the most distant groups and clusters. Large catalogs of low-mass systems together with the progress on the measurement of the thermal SZ power spectrum will open a new window into groups.

In the optical and near-infrared bands, space missions ({\it Euclid}\footnote{https://www.euclid-ec.org/} -from 2022- and {\it Nancy Grace Roman Space Telescope}\footnote{https://roman.gsfc.nasa.gov/} - formerly {\it WFIRST}; launch date: 2025) and ground based missions ({\it Vera Rubin Observatory}\footnote{https://www.lsst.org} and the 4-metre Multi-Object Spectroscopic Telescope, {\em 4MOST}\footnote{https://www.4most.eu/cms/}) will map the large scale structures over more than 15,000 deg$^2$, extending
the current catalogs of systems with M$_{\rm 500c} > 5 \times 10^{13}$M$_{\odot}$ (see, e.g., results from DES\footnote{https://www.darkenergysurvey.org/} in \citealt{DESY1cosmo}) by orders of magnitude, in particular at high ($z>1$) redshifts (e.g., \citealt{sartoris16}).  Of particular interest for the measurement of the velocity dispersions, is the {\em WFIRST} and {\em 4MOST} observatories.  The {\em 4MOST} observatory has been designed as a survey instrument at the forefront, with plans underway to combine the power of {\em 4MOST} with {\em eROSITA} (\citealt{finoguenov19}) in order to provide dynamical mass estimates for $\sim$10000 clusters at redshift z$<$0.6 and masses $>10^{14}$ M$_{\odot}$. $4MOST$ will also provide spectroscopic confirmation of $eROSITA$ detected groups at redshifts $<$0.2 down to a mass limit of 10$^{13}$ M$_{\odot}$.  Additionally, the wide area vista extragalactic survey (WAVES, \citealt{driver19})  being planned using $4MOST$, is aiming to perform the WAVES-Wide and WAVES-Deep surveys, allowing for the construction of optically selected groups catalogs.  The WAVES-WIDE(-DEEP) surveys aiming to cover an area of $\sim$70 (1200) deg$^{2}$, identifying $\sim$50000 (20000) dark matter halos down to a mass of $10^{14}$ ($10^{11}$) M$_{\odot}$ and out to a redshift of $z_{\rm phot} \lesssim$0.2 (0.8). Similarly to X-ray selected objects, optically-selected groups are physically heterogenous systems (e.g., see the dynamical analysis by \citet{2021ApJ...911..105Z} for a sample of compact groups). However, it is possible that the physical processes at work in the IGrM of optically-- and X-ray--selected systems are different.  Thus, the comparison of group samples selected via distinct methods can shed light on these physical phenomena.

\section{Final remarks}\label{sect:remarks}

Bridging the gap in mass between field galaxies and massive clusters, galaxy groups are key systems to make progress in our understanding of structure formation and evolution. Thanks to the current generation of X-ray satellites, together with dedicated hydrodynamical simulations, there have been significant improvement in our comprehension of the interplay between the hot ambient gas, radiative cooling and feedback due to, e.g., AGN activity and SNe winds, in particular in the central regions. 
Indeed, the thermodynamic  structure  of galaxy groups is more complex than in massive galaxy clusters, with the physics associated to non gravitational processes playing a significant role in shaping their general properties. The scaling relations capture the result of the various (thermal and non thermal) processes and show that galaxy groups are not simply the scaled-down versions of rich clusters. 
Thanks to the enlarged catalogs of low-mass systems that the current (and upcoming) wide surveys at X-ray, millimeter, and optical wavelengths will provide, such scaling relations can be measured with very high precision. The comparison between results obtained from differently selected samples will shed light on the 
intrinsic properties of the groups' population.

Due to the complexity of the X-ray emitting processes in the low-temperature regime, and of
how AGN heating impacts the general properties of the core of poor systems, 
the interpretation will depend on the specific choices of the individual analyses. 
For instance, many X-ray studies on groups (and clusters) provide integrated measurements within a certain aperture. However, the definition of such aperture is often quite different, and the comparison between the various works is not always straightforward. 
In the future, it is desirable to provide the global properties using an unified definition of the regions, that are efficiently accessible from observations. This will ease the comparison between different observational and theoretical studies, improving our understanding of the physical processes at work in the complex group regime. Of course, in each study there are good reasons to use a specific energy band or definition of region of interest. However, regardless of the  choices made in each paper to reach specific goals, we suggest to also provide, whenever possible, both global and core-excised properties within R$_{500}$. Although there are evidences that the cool-core radius does not scale uniformly with the virial radius, we think that the common choice of excising $r<$0.15R$_{500}$ is a good starting point. For the rest-frame luminosities, we have shown that the 0.5-2 keV band is less sensitive to the choice of the abundance table and is easily accessible for all the current and future facilities (differently from the 0.1-2.4 keV band which extend to a regime where, for instance, {\it Chandra} and {\it XMM-Newton} are not well calibrated and also the choice of the abundance table start to play a role as discussed in Section \ref{selfsimilar}). However, to ease the comparison with the literature it is useful to provide also the rest-frame bolometric and 0.1-2.4 keV band luminosities. Finally, until R$_{500}$ will be routinely achieved for most of the systems in the low-mass regime,  we also suggest to provide the properties at R$_{2500}$ (i.e., $\sim$0.5R$_{500}$). 
Of course, there are further complications (e.g., the impact on the temperature of using a certain abundance table or spectral code, \citealt{LL15}; the choice of the column density, \citealt{LL19}; the fitting technique, \citealt{balestra07}) which play a relevant role in the low-mass (but not only) systems. Nonetheless, starting to set standard definitions will definitely help the analysis in this critical regime.

\vspace{6pt} 



\authorcontributions{Conceptualization, L.L, S.E., M.G., and 
P.A.G.; resources, L.L, S.E., M.G., and P.A.G.; data curation, L.L., S.E., M.G., and P.A.G.;  writing-original draft preparation, L.L, S.E., M.G., and P.A.G.; writing-review and editing, L.L, S.E., M.G., and P.A.G.; visualization, L.L, S.E., M.G., and P.A.G.; supervision,
L.L; project administration, L.L. 
All authors have read and agreed to the published version of the manuscript.}

\funding{L.L. and S.E. acknowledge financial contribution from the contracts ASI-INAF Athena 2015-046-R.0, ASI-INAF Athena 2019-27-HH.0, ``Attivit\`a di Studio per la comunit\`a scientifica di Astrofisica delle Alte Energie e Fisica Astroparticellare" (Accordo Attuativo ASI-INAF n. 2017-14-H.0), and from INAF ``Call per interventi aggiuntivi a sostegno della ricerca di main stream di INAF".  M.G. acknowledges partial support by NASA Chandra GO8-19104X/GO9-20114X and HST GO-15890.020-A grants.  P.A.G. acknowledges support from the UK Science and Technology Facilities Council via grants ST/P000525/1 and ST/T000473/1.}

\acknowledgments{
The authors thank the anonymous referees for useful comments and suggestions that helped to improve and clarify the presentation of this work. We thank M. Sun  for providing the luminosity values for the group sample analysed in his work. We also thank MNRAS and the AAS, together with the authors of the corresponding publications, for granting permission to use images published in their journals.
}

\conflictsofinterest{The authors declare no conflict of interest.} 

\appendixtitles{no} 



\reftitle{References}


\externalbibliography{yes}
\bibliographystyle{Definitions/aa}
\bibliography{SLbib,biblio2}



\end{document}